\documentclass{aa}  
\usepackage{natbib}
\usepackage{graphicx}
\usepackage{txfonts}
\usepackage{aalongtable}
\newcommand{\hm}{H$_{2}$}

\newcommand{\dm}{D$_{2}$}

\begin{document}

\title{Water  formation on bare grains: When the chemistry on dust impacts interstellar gas.}
\titlerunning{water on bare grains}
\author{S. Cazaux\inst{1}, V. Cobut\inst{2},   M. Marseille\inst{3}, M. Spaans\inst{1}, P. Caselli\inst{4}} 
\offprints{cazaux@astro.rug.nl}
\institute{Kapteyn Astronomical Institute, PO box 800, 9700AV Groningen, The Netherlands\\
\and
LERMA, UMR 8112 du CNRS, Observatoire de Paris et Universit\'e de Cergy Pontoise, 5 Mail Gay-Lussac, F-95031 Cergy-Pontoise Cedex, France
\and
SRON, National Institute for Space Research, PO Box 800, 9700 AV Groningen, The Netherlands
\and 
School of Physics and Astronomy, University of Leeds, LS2 9JT, Leeds, UK}

\date{Received ; accepted }

\abstract {Water together with O$_2$ are important gas phase ingredients to cool dense gas in order to form stars. On dust grains, H$_2$O is an important constituent of the icy mantle in which a complex chemistry is taking place, as revealed by hot core observations. The formation of water can occur on dust grain surfaces, and can impact gas phase composition. }{The formation of molecules such as OH, H$_2$O, HO$_2$, H$_2$O$_2$, as well as their deuterated forms and O$_2$ and O$_3$ is studied in order to assess how the chemistry varies in different astrophysical environments, and how the gas phase is affected by grain surface chemistry.}{We use Monte Carlo simulations to follow the formation of molecules on bare grains as well as the fraction of molecules released into the gas phase.  We consider a surface reaction network, based on gas phase reactions, as well as UV photo-dissociation of the chemical species.}{\rm We show that grain surface chemistry has a strong impact on gas phase chemistry, and that this chemistry is very different for different dust grain temperatures. \rm Low temperatures favor hydrogenation, while higher temperatures favor oxygenation. Also, UV photons dissociate the molecules on the surface, that can reform subsequently. The formation-destruction cycle increases the amount of species released into the gas phase. We also determine the time scales to form ices in diffuse and dense clouds, and show that ices are formed only in shielded environments, as supported by observations.}{}


\maketitle

\section{Introduction}
There is water everywhere in the cosmos. In cold and dense regions of the interstellar medium, dust grains become covered by icy mantles mainly made of water (\citealt{williams1992}; \citealt{gibb2004}). In the star formation process, the ices that were retained onto dust grains evaporate and drive a very rich chemistry (\citealt{caselli1993}; \citealt{cazaux2003}). At a latter stage, when planets are formed, the water still trapped in solid form is carried by asteroids and comets, can be brought to the different planets, giving birth, on our own planet, to oceans.

Gas-phase chemical theory predicts that H$_2$O and O$_2$, along with O and CO, constitute the gas-phase reservoirs of elemental oxygen in molecular gas that is well shielded from ultraviolet radiation (e.g., Millar 1990). As a consequence, water together with O$_2$ can be important coolants for dense gas (\citealt{goldsmith1978}; \citealt{hollenbach1988}; \citealt{Neufeld1995}).

The formation and constitution of icy mantles has been studied by several authors (\citealt{tielens1982}, \citealt{cuppen2007}). Observations revealed that icy mantles are sensitive to hard radiation, and therefore could only exist in shielded environments (\citealt{whittet2001}, \citealt{papoular2005}). Ices are supposed to be the result of gas phase species that freeze out onto dust grains, that were previously formed on the grain or in the gas phase. \rm{Because the binding energy of water on bare carbonaceous grains is lower than water on ice, the first layer of ice is essential to form the icy mantle (\citealt{papoular2005}) and could have an important impact on the snow line position (Marseille \& Cazaux submitted). Note that for water sticking on bare silicate grains this could be different ( \citealt{goumans2009}, \citealt{schiff1973})}\rm

Because of the inefficiency of gas phase routes, grain surface chemistry is required to explain the observed abundances of water in molecular clouds (\citealt{d'Hendecourt1985}; \citealt{hasegawa1992}).  H$_2$O abundances can range from $<$ 10$^{-8}$ in the coldest clouds, that can be explained by freeze-out of water on dust, to $>$ 10$^{-4}$ in warm gas and shocks, due to ice evaporation or sputtering and exothermic gas phase reactions (\citealt{bergin1998}, 2000, \citealt{hollenbach2009}).

In this work we will investigate the formation of molecules such as OH, H$_2$O, H$_2$O$_2$ and their deuterated forms in different astrophysical environments. In particular, we will concentrate on grain surface chemistry, and the amounts of species that this process releases into the gas phase. \rm{The interactions between species and carbon surfaces (graphite or amorphous carbon) has received a lot of attention (\citealt{ghio1980,pirronello1997,jeloaica1999,sha2002,bergeron2008}}). Theoretical and experimental studies of the interactions between chemical species and these types of surfaces represent a considerable number of constraints for our model. Silicates, on the other hand, are more complex surfaces that are more difficult to treat theoretically, resulting in fewer constraints for the interactions between species and surface. However, recent studies from \cite{goumans2009} revealed that the formation of water differs strongly as it is formed on olivine or on carbon grains.  On olivine surfaces, the formation of water occurs via species strongly bound to the surface that can dissipate the excess energy in an efficient way via phonons. This  allows most of the newly formed species to stay on the surface. Therefore, its seems that the formation of species on olivine surfaces would not impact the gas phase significantly. In this study, because we want to assess how the chemistry on dust grains can impact interstellar gas, we exclusively follow the chemistry that occurs on carbon grains. To do so,  we use Monte Carlo simulations to describe the formation of chemical species on such surfaces. Also, we consider several environments with different physical conditions, and establish how chemical species form on dust as well as their subsequent released in the gas phase. Finally, we discuss the formation of icy mantles in atomic and molecular clouds, and show that our model reproduces the fact that grains are covered by ices only in shielded environments.\rm

\section{Grain surface chemistry}
\subsection{Formation of water and its deuterated forms with Monte Carlo simulations}
In a previous paper (\citealt{cazaux2008}), we discussed the formation of \hm, HD and \dm\ on surfaces that are typical to the interstellar medium. We found that the formation of molecules depends on the binding energy of atoms with the surface and  on the barrier that atoms from the gas phase have to cross to become strongly bound to the surface. Indeed, there are \rm{two types of interactions}\rm\ between the atoms and the surface: a weak one, called physisorption (van der Waals interaction), and a strong one, called chemisorption (covalent bound). Atoms from the gas phase can access easily the physisorbed sites and become physisorbed atoms. These weakly bound atoms can scout the surface at very low dust temperatures, meeting each other to form molecules. At higher dust temperature, the physisorbed atoms evaporate and the formation of molecules is insured by the contribution of chemisorbed atoms. The atoms from the gas phase can also enter in a chemisorbed site (and have to cross a barrier to do so), and can meet an already chemisorbed atom. This process will allow molecules to form for a wide range of dust grain temperatures (up to few hundreds of kelvins). 

\rm{We use a  step by step Monte Carlo simulations  to follow the chemistry occurring on dust grains. The dust grains are divided into square lattices of 100$\times$100 adsorption sites. Each site on the grid corresponds to a physisorbed site and a chemisorbed site, so that the grain can be seen as 2 superimposed grids. Species that are coming from the gas phase arrive at a random time and location on the dust surface. This arrival time depends on the rate at which gas species collide with the grain. This rate of accretion can be written as:
\begin{equation}
R_{acc} = n_X v_X  \sigma  S. 
\end{equation}
 where n$_X$ and v$_X$ are the densities and velocities of the species X, $\sigma$ is the cross section of the dust particule and $S$ is the sticking coefficient of the species with the dust. We consider S=1, meaning that when a species arrives at a point of the grid, it can become physisorbed, or, if its chemisorption states exists and its energy is high enough to cross the barrier against chemisorption, becomes chemisorbed. \cite{jones1985} show that the sticking coefficient depends on the proportion physisorbed/chemisorbed sites on the surface, and that for a gas at 100K, most of the hydrogen atoms populate the chemisorbed sites. In our model, on the other hand, because of  the high barrier to access chemisorbed sites, H atoms mostly arrive from the gas phase in physisorbed sites.  In this sense, we overestimate the sticking coefficient since all atoms that arrive on the grain stick. 
 
 The species that are present on the surface can go back to the gas phase because they evaporate. This evaporation rate of the species i can be written as:
 \begin{equation}
R_{evap(i)} = \nu_i \times \exp{(-\frac{E_i}{k_BT})} 
\end{equation}
where E$_i$ is the binding energy of the considered species i, as reported in Table~\ref{table1}.
 
The species that arrive at a location on the surface can move randomly, \rm{on the surface through tunneling effects and thermal hopping. The diffusion rates R$_{ij}$ for an atom to go from a site i (physisorbed or chemisorbed) to a site j (physisorbed or chemisorbed) can be written as:
\begin{equation}
R_{ij} = \nu_i \times P_{ij}. 
\end{equation}
where $\nu_i$ is the oscillation factor  of the atom in the site i (which is of 10$^{12}$ s$^{-1}$ for physisorbed atoms), and P$_{ij}$ is the probability for the atom to go from a site i to j by tunneling effect or thermal hopping, as described in \cite{cazaux2004}. When thermal hopping dominates, which is usually the case for species going from a physisorbed sites to another physisorbed site, or for high temperatures, this probability can be written as $P_{ij}=\exp({-\frac{E_{ij}}{k_B T}})$,  where $E_{ij}$ is the energy of the barrier between the site i and j, and T the temperature of the surface. The hopping barriers to go from a physisorbed site to another physisorbed site are taken as  2/3 of the binding energy (E$_i$, listed in table~\ref{table1}) of the species on the surface, with a width of 3\AA, which is what we derived for the case of \hm\  formation of carbonaceous and silicate surfaces (\citealt{cazaux2004}). The barrier for a physisorbed atom to become chemisorbed is discussed in section 2.2.2.

 Species present on the surface can also receive a UV photon and become dissociated with rates reported in section 2.4. Once 2 species meet in the same site, they can form a new species if the activation barrier for the formation can be overcome. \rm\ The different reactions that can occur on the dust grains as well as their activation barriers $E_a$ are listed in Table~\ref{table1}. \rm{When the barrier for a reaction has not been studied on the surface, we assume that the reactions are similar to gas phase reactions and use the NIST database (\citealt{manion2008}). However, recent studies showed that reactions that have a barrier in the gas phase can be barrierless on the dust surface (\citealt{ioppolo2008}).\rm\ Each time a reaction with a barrier occurs, we calculate the rate for crossing the barrier, which is $R_{E_a}=\nu\times P_{E_a}$, where $\nu$ is the oscillation factor of the considered species, and  $P_{E_a}$ is the probability for a chemical species thermalized on the grain at a temperature $T_{\rm{dust}}$ to cross the activation barrier $E_a$. We assume here that these barriers have a width of 1\AA. We compare the rate of crossing the barrier $R_{E_a}$ to the rate for the chemical species to move out of the site R$_{ij}$. If the rate of leaving the site is lower, then the reaction will occur, if not, the species will leave the site \rm{and go to a neighboring site. In this study we assume that species that remain on the grain are thermalized, and therefore that there is no preferential direction for the dissociated products that stay on the surface. We therefore consider that the probabilities to visit neighboring sites are equal.}\rm

Once a chemical species present on the dust moves, or adsorbs a photon, or meets another species, the next event that will occur to this species is determined as well as its time of occurrence. Therefore, the events that concern every species on the dust are ordered by time of occurrence, and for each event that occurs, a next event for the concerned species is determined. The chemical species considered in our model are the following: H, D, O, OH, OD, O$_2$, H$_2$O, HDO, D$_2$O, O$_3$, HO$_2$, H$_2$O$_2$, DO$_2$, HDO$_2$ and D$_2$O$_2$. Apart from H and D, that can also become chemisorbed as discussed in the next section, all the other species only fill the physisorbed sites. 

With the inclusion of deuterium, different products can be formed for the same reaction. We report the branching ratios of these reactions in table~\ref{table1}. For example, if we consider the reaction HD + O, the HD molecule has to be dissociated, and the oxygen reacts either with H or D. \rm{For this reaction, the branching ratio are strongly dependent on the temperature. At low temperatures,  tunneling dominates the reaction and  HD + O gives mostly OH + D (with a rate k$_{HD}$) and very few OD + H (k$_{DH}$). At 300K the ratio $\frac{k_{HD}}{k_{DH}}$ is of the order of 15 (\citealt{robie1987} ), and strongly increases as temperature decreases. Therefore, we assume that at low temperatures, the branching ratio k$_{DH}$ is negligible compared to k$_{HD}$.} \rm  For the case of H$_2$O$_2$ + D, the reaction of one of the H atoms with a D atom, which gives HD + HO$_2$,  then breaks the O=O bound, and finally adds the free oxygen to HD. The final products are therefore: OH + HDO (\citealt{albers1971}). \rm For the other reactions, when no branching ratios have been reported (in the NIST database) we assume equal branching ratios.\rm

\subsection{Binding energies}

\subsubsection{Physisorption}
The binding energies of the different species in physisorbed sites on carbon surfaces have been discussed by \cite{cuppen2007} and references therein. As discussed by these authors, the binding energies of species on the surface are poorly known. On amorphous carbon grains, it seems that only H and \hm\ physisorption energies have been determined (\citealt{pirronello1997}, \citealt{katz1999}). For most of the other binding energies, we refer to graphitic surfaces. H atoms are bound to graphitic surfaces with an energy of 550~K (\citealt{bergeron2008} and references therein, \citealt{ghio1980}), which is similar to the amorphous carbon one. The physisorption energy of molecular hydrogen on carbon nanoparticles (graphitic platelets) and polycyclic aromatic hydrocarbons (PAHs) is determined to be attractive between 420 and 860~K, depending on the orientation of \hm\ and on the particle size. These energies are also dependent on the temperature of the surface (\citealt{heine2004}). Therefore, we use a binding energy of 600~K, as it has been determined for graphene surfaces (\citealt{akai2003}).  \cite{bergeron2008}  found that oxygen can be physisorbed on graphitic surfaces with energies of 1400~K, if the atom is placed on a top site (above a C atom) or on a bridge site (between two C atoms). This value is quite high compared to the value of 800~K used in \cite{cuppen2007}, initially taken from \cite{tielens1982}. In a recent article, \cite{lee2009} calculated the binding energy of ozone on graphene. Ozone molecules adsorb on graphene basal plane with binding energy of 2900~K, and the physisorbed molecule can chemically react with graphene to form an epoxide group and an oxygen molecule. The activation energy barrier from physisorption to chemisorption is very high (8000~K), and we will therefore neglect the fact that ozone molecules would overcome this barrier, and be dissociated. For the other species, we consider similar energies as in \cite{cuppen2007}. The energies of physisorption of the different species are reported in Table~\ref{table2}. 

\subsubsection{Chemisorption}
Hydrogen atoms can chemisorb on graphite surfaces, but this process has an associated activation energy of 0.2 eV (\citealt{hornekaer2006}, \citealt{sha2002}, \citealt{jeloaica1999}). Recent studies show that once a first H atom is chemisorbed on a graphite surface, a second atom can chemisorb on the same benzene, in the site opposite to this first atom (para site), without a barrier (\citealt{hornekaer2006}, \citealt{rougeau2006}) and a third atom will form a molecule with one of these two adsorbed atoms without a barrier (\citealt{bachellerie2007}). This process is important to form \hm\ efficiently at high dust temperatures. This process will be added to our simulations in a coming paper (Cazaux \& Morisset, in prep.). \rm{For now, we only consider that H atoms can become chemisorbed (with an energy of 0.7 eV) by crossing a barrier of 0.2 eV by tunneling effects (low temperatures) or thermal hopping (high temperatures), and approximate the barriers as being squared. Nevertheless, DFT calculations show that these barriers are far from being square resulting in higher rates for H atoms to become chemisorbed. This will be discussed in a future work (Cazaux \& Morisset in prep.)} \rm If the process discussed by \cite{bachellerie2007} would be taken into account, once a first atom is present in a chemisorbed site, there is a barrier-less way to form and desorb a \hm\ molecule. In this study, we are thus underestimating the rate of \hm\ formation. This can have a small effect on the formation of molecules in diffuse clouds, which are mostly atomic environments, but has no effect for molecular environments, since the reactions to form molecules involve \hm.

Theoretical and experimental studies show that oxygen atoms chemisorb preferably above the middle of the C-C bond, creating a stable epoxide-like structure. Oxygen has adsorption energies which are $\sim$ 2.3-2.4~eV, with a barrier against chemisorption of the order of 0.15-0.2~eV, depending on whether the O chemisorbs above a C or in the middle of the C-C bond (\citealt{fromherz1993}). \rm In the singlet state, oxygen atoms can be chemisorbed with an energy of 2.5~eV (\citealt{jelea2004}), and can react with H atoms to form OH radicals through an Eley-Rideal mechanism with an activation barrier that varies from 0.17 eV to 0.29 eV.  However, oxygen is mostly found in its triplet form in the ISM. Recent studies by \cite{bergeron2008} show that  the chemisorption of oxygen in a triplet state has an associated barrier of 0.2~eV, with a ''meta-stable'' chemisorbed state, meaning that the well lies above the energy level for $^3$O + graphite at infinite separation. Therefore, only for high temperature gas (2000~K) the chemisorbed state of oxygen (triplet) can be populated. In this work, because oxygen is mostly found under the triplet form in the ISM, we will consider that oxygen atoms cannot be chemisorbed.}\rm
 
\subsection{Desorption upon formation}
Once a reaction occurs on the surface, there is a certain probability that the product of the reaction desorbs into the gas phase. This process depends on the energy released upon formation (enthalpy of reaction) and on the binding energy of the final product. \rm{The desorption upon formation concerns only physisorbed species that react with each other to form a product that is also physisorbed. Because of the excess energy contained in the molecule, when reactions have high enthalpies, this physisorbed product can be ejected into the gas phase.This process has been discussed in a recent work by \cite{garrod2007}. These authors quantify the probability of desorption using Rice-Ramsperger-Kessel (RRK) theory. However, this theory is designed for large molecules, and we expect a different behavior for diatomic molecules.} \rm With this method, they derive a probability $P$ that for a certain reaction, a certain energy $E$, higher than the binding energy of the product ($E \ge E_B$), is found in the bond. This probability can be written as:
\begin{equation}
P=\left[ 1- \frac{E_B}{E_{\rm{reac}}} \right]^{s-1},
\end{equation}
where the energy released during the reaction is $E_{\rm{reac}}$, and $s$ is the number of atoms in the molecule which is produced.
Once this probability is known, these authors use a method that depends on the energy loss on the surface, and that calculates the fraction of newly formed molecules that would desorb from the surface upon formation. This fraction is written as:
\begin{equation}
P=\frac{\nu P}{\nu_s+\nu P} = \frac{a P}{1+a P}
\end{equation}
where $\nu_s$ is the rate at which the total energy is lost to the surface, $\nu$ is the oscillation factor of the molecule in the potential well, and $a=\frac{\nu}{\nu_s}$. We can speculate that the energy loss will depend on the porosity of the surface, meaning that a newly formed species can loose more efficiently its energy on a porous surface, since it will encounter more obstacles to bump into. Because water ices are supposed to be more porous than bare surfaces, molecules produced on bare grains should have higher probabilities to desorb directly into the gas phase upon formation. \cite{garrod2007} use the data of \cite{kroes2005} to determine the rate $\nu_s$, and derive a factor $a$ = 0.012. In the case of bare grains, on the other hand, \cite{katz1999}, \cite{cazaux2004}, \cite{cuppen2005}, derive from the experiments of \cite{pirronello1997} that the fraction of \hm\  that directly desorbs into the gas phase upon formation is about 70~$\%$ for silicates and 60~$\%$ for amorphous carbon grains. With the theory developed in Garrod et al. (2007), we obtain 1~$\%$, which is far from what has been derived for \hm. If we consider that the energy loss is much less efficient in the case of bare grains, meaning that the factor $a$ has to be a few orders of magnitude higher than the 0.012 determined by Garrod et al. (2007), we obtain a very similar fraction of  desorption \rm{upon formation}\rm\ for \hm\ and H$_2$O (if $a$ = 0.7, we obtain 40~$\%$ and 39~$\%$ for the desorption upon formation of \hm\ and H$_2$O, respectively). In the discussion of \cite{cuppen2007}, it seems that the fraction of water that desorbs upon formation is quite small (\citealt{kroes2005}). Actually, this study deals with water ice, but for non-porous surfaces such as bare grains this fraction is different. 

Recent studies from \cite{bergeron2008}, show that once \rm{physisorbed H and O atoms recombine on graphite surface}\rm, the nascent molecules can either directly desorb, if the translational energy is large enough, or the molecule is trapped in a quasi stationary state where it oscillates in the physisorption well with a signiÞcant amount of vibrational and rotational energy. As the H atom repeatedly collide with O, OH escapes from the surface (the so-called complex mechanism). Also, \cite{bachellerie2007} show that \hm\ molecules mostly desorb from the surface upon formation. This study shows that on a graphene surface at low temperatures, only 3$\%$ of the newly formed \hm\ molecules stay on the surface (they consider species staying more than 12~ps as trapped). Therefore, we can assume that a significant fraction of newly formed species can desorb directly into the gas phase upon formation.

We therefore will use an approximation that has been derived by Vasta et al. (2005, unpublished), where the desorption of \hm\ is high, and of H$_2$O is low, depending on the binding energy of the final product and the enthalpy of reaction. This approximation, which reproduces the fraction of \hm\ that desorbs upon formation on silicates and amorphous carbon, also keeps this fraction small for water, and  is more adequate for our purpose of modeling the formation of chemical species on bare grains.  
\begin{equation}
P=\frac{E_{\rm{reac}}}{4.5\times11600}\times(0.756-3\times 10^{-4}E_{\rm{B}}).
\end{equation}
\rm This empirical formula  reproduces that 70~\% (60~\%) of \hm\ molecules desorb upon formation on silicate (amorphous carbon) surfaces, where \hm\  molecules have a binding energy of 300~K (500~K). The fraction of molecules that desorbs upon formation linearly scales with the enthalpy of reaction.\rm In our model, 60~$\%$ of the newly formed \hm\ molecules desorb upon formation, whereas this fraction is 36~$\%$ for the reaction H + O $\rightarrow$ OH and 15~$\%$ for OH + H $\rightarrow$ H$_2$O. This fraction depends on the route to form the chemical species, since it becomes only 4~$\%$ when OH is formed through O + \hm\, and 0.8~$\%$ when H$_2$O is formed through OH + \hm. Therefore, the fraction of species released into the gas phase will vary strongly if the species are formed in atomic or molecular environments.  

\subsection{Photo-dissociation and photo-desorption}
Interstellar dust grains can be present in environments that are subject to radiation. Stars in the neighborhood can emit FUV photons that impinge on the interstellar dust grains. The flux of FUV photons can be written as follows (Hollenbach et al. 2008):
\begin{equation}
F_{\rm{FUV}}=G_0 F_0 \exp^{-1.8 \rm{A_V}}
\end{equation}
with $F_0\sim$10$^8$ photons cm$^{-2}$ s$^{-1}$, the approximate local interstellar flux between 6 and 13.6 eV,  $G_0$ a scaling factor for which the value 1 corresponds to the Milky Way interstellar radiation field in the FUV band, and $A_V$, the visual extinction. 

Once a UV photon arrives on a species, this species can be photo-dissociated. The products of the dissociation can be released directly into the gas phase, or one or both fragments can be trapped on the surface. \cite{andersson2008} calculated the outcomes of the photodissociation of water molecules in water ices. Once a water molecule is broken into OH and H in the first monolayer of ice, several processes can occur with different probabilities if the ices are crystalline or amorphous (\citealt{andersson2006}):  a) H atoms are released into the gas phase and OH molecules are trapped (amorphous: 92~$\%$, crystalline: 70~$\%$); b) H and OH are trapped (amorphous: 5~$\%$, crystalline: 14~$\%$); c) OH and H desorb (amorphous:  2~$\%$ and crystalline:  6~$\%$); d) OH and H recombine and form H$_2$O which can be trapped (amorphous: 1~$\%$, crystalline:  0.05~$\%$) or desorb (amorphous:0.7~$\%$, crystalline: 1~$\%$).  These results are different from crystalline and amorphous ices. In particular,  H atoms due to dissociations are less trapped in crystalline ice than amorphous ice.  According to these authors, the yield of desorption of water molecules is rather small (0.2~$\%$) for both types of ices, as confirmed by experiments  from \cite{oberg2008}. 

In our simulations, we consider that an  UV photon absorbed by a water molecule can lead to desorption of OH and H with a probability of 2~$\%$. For the rest, we consider that the products stay on the grain surface as OH and H. Because we consider bare surfaces, OH and H have different mobilities and therefore the probability for the reaction and reformation of the water molecule should be different than on ices. We also consider that H atoms stay on the grain upon dissociation.  \rm{Because of our lack of knowledge concerning the direct photo-desorption of species other than water, we only consider the desorption of water as OH and H into the gas phase. For the other species, we consider that there is no photo-desorption, but that photo-dissociation is possible, as discussed in the next paragraph. This seems to be a correct approximation, since, as discussed by  \cite{andersson2008}, photo-dissociation (processes a and b discussed above) is more important than direct photo-desorption (process c).}\rm

In this study, we consider that only the photons that arrive directly on the species can photo-dissociate them, meaning that the cross section of reaction is similar than the one in the gas phase. Table~\ref{phot2} presents the different photodissociation reactions and their associated photo-dissociation rates. Also, we consider that 2$\%$ of the products from a photo-dissociation reaction would directly desorb into the gas phase, while 98$\%$ of the same products remain on the grain surface. We do not consider here the possibility of one product staying and the other one desorbing. We also take into account cosmic-ray-induced photons. These photons can interact with species present on the grain surface, and dissociate them with rates similar to the gas phase ones. These rates are also reported in Table~\ref{phot2}. \rm{In our simulations, we consider that the products of the photodissociation are directly thermalized on the surface, meaning that they loose their excess energy in very short time scales. The 2 products of the dissociation  are placed in neighboring sites. \rm

\subsection{Water clusters}
If a water molecule is located in a site next to another water molecule,  a dimer can be created, making the total binding energy larger. \cite{bolina2005} and \cite{brown2007} performed Temperature Programmed Desorption (TPD) experiments for the desorption of pure H$_2$O ice on a graphite surface (highly ordered pyrolytic graphite). The experimental data showed that with increasing coverage, water molecules desorb at higher temperature.  Water molecules are arranged in clusters, which have an increasing binding energy with increasing size. Therefore, as the coverage increases, the cluster becomes more important, and a higher surface temperature is needed to desorb the molecules. Ab initio calculations from \cite{lin2005} and \cite{gonzalez2007}, presented the adsorption energies of a single water molecule, but also a dimer, trimer, until the hexamer. These binding energies increase with the number of water molecules that are present in the cluster, but only if the cluster is in 2 dimensions. \cite{lin2005} show that the binding energy of a 3D cluster is mainly dependent on the number of water molecules which are close to the graphite surface. In this study, because we consider grain surfaces at low temperatures ($\le$30~K), we do not take into account the formation of water clusters, since any newly formed water molecule that is not ejected upon formation stays on the surface. Water clusters and their impact on the freeze-out of water mantles and the snow line in proto-stellar environments is treated in another work (Marseille \& Cazaux Submitted).

\section{Results}
\subsection{Test case:  the effect of the dust temperature and UV photons}
We calculate the number of molecules formed on the dust grain and released into the gas phase as a function of time for different dust grain temperatures. Note that we only consider grain surface chemistry, and that the gas phase reactions and abundances are not treated in this paper. We start our analysis with a high O/H and D/H ratio of 0.1, and $n_{\rm{H}}$ = 10$^3$~$cm^{-3}$, in order to understand how grain temperature influences the chemistry occurring on interstellar grains. We also add a UV radiation field to assess the changes in grain surface chemistry for the presence of photons. \rm{In this section we consider parameters that are not representative of any astrophysical environment in order to understand  how these parameters influence the chemistry on dust and its impact on gas phase.}\rm

The population of the chemical species present on the dust grain and released into the gas phase, in absence of UV radiations, are presented in Fig.~\ref{temp1}.  When looking at the fraction of the species covering the grain (with monolayer  equal to 1 corresponding to a surface that is completely covered), it appears that the chemistry occurring on the dust grain is very different if the grain has a temperature of 10~K or 30~K. At 10~K, the atoms and molecules do not evaporate into the gas phase, and the H atoms (and D atoms) are the only species mobile on the surface. Therefore, the dominating process is hydrogenation and most of the oxygen is found in the form of H$_2$O and HDO.  On the grain surface, as shown in Fig.~\ref{temp1}, top left panel, H$_2$O and HDO are the dominant species. Each time an OH species is formed and stays on the dust grain, a H (D) atom already present on the surface finds this OH (OD) radical and associates to form H$_2$O (HDO). Therefore, the successive hydrogenations are very efficient, and O converts rapidly to OH and H$_2$O.  The species released into the gas phase, represented in fig.~\ref{temp1}, top right panel, are identical to the ones formed on the grain: $\sim$30~$\%$ of the oxygen coming on the grain is released in the gas phase as OH, and $\sim$5~$\%$ as H$_2$O. 
On 30~K grains, on the other hand, the H and D atoms evaporate, while oxygen atoms, more strongly physisorbed to the surface, stay and can encounter each other. Therefore, lots of oxygen bearing species are formed that contain more than one oxygen atom. The grain surface composition is rich in H$_2$O and HDO (Fig.~\ref{temp1}, bottom left panel). Lots of chemical species are only transitory, such as OH, OD, HO$_2$ etc. and turn very fast into H$_2$O, which explains their absence from the grain surface. The species released into the gas phase are more complex than grain surface species, as shown in Fig.~\ref{temp1}, bottom right panel. Oxygen atoms coming on the grain are released into the gas phase in the form of OH (20~$\%$), O$_2$ (10~$\%$), H$_2$O (5~$\%$), OD (2~$\%$) and H$_2$O$_2$ (1~$\%$). These species are released into the gas upon formation.  At these temperature ranges, this process is the most important to populate the gas phase with species formed on dust since evaporation is negligible. 

\rm{When dust grains are subject to strong UV radiation fields, the species present on their surface can be photo-dissociated. If we consider 10~K and 30~K dust grains in an environment subject to a UV radiation field of $G_0 = 10^3$, and at an extinction of $A_V$ = 1 mag, the molecules on the grain surface are photo-dissociated very fast (Fig.~\ref{temp2}, left panels). Each time a molecule forms and stays on the surface, it can be destroyed by an incident UV photon. Because of the high fraction of atoms on the surface, OH radicals are transitory species, and turn into H$_2$O very fast. The fraction of H$_2$O that covers the surface reaches an equilibrium between photo-dissociation into OH and H and reformation. The amount of water that is released in the gas phase upon formation is enhanced by the presence of the UV radiation field (Fig.~\ref{temp2},  right panels). Without UV field, this amount is on the order of $5\%$. This fraction becomes much higher $\sim$12-15 $\%$ in the presence of a strong radiation field. This is due to the fact that each water molecule that forms has a probability to be released in the gas, but if this molecule stays, it can be dissociated and reform with the same probability of being desorbed. In this sense, the water molecules follow a cycle formation-destruction that enhanced the probability of being released in the gas phase. Analytically, we find that the amount of water that directly desorbs into the gas upon formation should be $\sim$5.4$\%$(0.64$\times$ 0.15 = 0.054, amount of OH that stays on the surface upon formation times the amount of H$_2$O that desorbs upon formation). Now when a strong UV radiation field is present, the molecules can be dissociated in products that can meet, and reform the same molecules. The probability that the H atom coming from the dissociation meet the other product OH to reform H$_2$O is determined in the Appendix A. We determined that this process, commonly called backdiffusion, has  a probability of 45$\%$ to occur, and therefore enhances the fraction of water released into the gas phase by a factor 2 ($\sim$ 10$\%$ see Appendix A). This value is slightly lower than the one we found with our simulations, meaning that backdiffusion is enhanced by grain size limitation that we did not consider. Our calculations over the backdiffusion in Appendix A differs from previous studies from  \cite{chang2005} who studied the H + H system and from \cite{krug2004} who concentrated on stepped surfaces . In our study, only H atoms can move on the surface of bare grains, and we concentrated on the probability for H atoms resulting from the dissociation of water molecules to visit the location of the other product of the dissociation. \rm

It seems obvious that low dust temperatures favor hydrogenation and deuteration of the species on the surface. For higher dust temperatures, hydrogen and deuterium, can evaporate into the gas phase because of their low binding energies, while oxygen atoms, with higher binding energy, become mobile on the surface without evaporating. The resulting chemistry  is oxygenation which leads to a chemistry rich in oxygen bearing molecules. 
UV photons can dissociate most of the species on the surface. \rm{The products of the dissociation can meet again and reform the original species. Because each new species formed has a probability to be ejected into the gas phase upon formation, the cycle formation-destruction-formation increases the total fraction of species released into the gas phase. }\rm

\section{From diffuse to dense clouds and PDRs}
In this section, we discuss the formation of molecules on grains in different environments. The parameters used in our simulations are based on \cite{snow2006} and are reported in Table~\ref{para}. \rm{For each environments, we performed at least 4 different simulations with different seeds, and we report the errors bars that represent 95$\%$ of confidence level}. 

\rm We first compute the chemistry occurring on dust grains in diffuse clouds, where the gas phase composition is mostly atomic. We assume a density of 100 atoms cm$^{-3}$, a visual extinction of 0.5~mag and an impinging UV radiation field of $G_0$=1. The species present on dust grains and released into the gas phase are reported in Fig.~\ref{env1}. In this environment, the process to form water is through the association of H and O atoms. The water formed on the grains can desorb or remain. The fraction of the incoming oxygen atoms that come and leave the grain as OH is $\sim 35~\%$, while it is $\sim 10~\%$ as H$_2$O. Indeed, an oxygen atom that arrives on a grain and meet a H atom, will form OH. Because of the energy released during this reaction, OH desorbs in 35~$\%$ of the cases. The fraction that remains on the surface (65~$\%$) can meet another H atom to form H$_2$O. Again, this reaction is exothermic and in 15~$\%$ of the cases, the product H$_2$O is ejected into the gas phase, making a total of 10~$\%$ of the O transforming into gas phase H$_2$O (0.65$\times$0.15 = 0.1). The water that remains on the surface builds up very slowly until it reaches 0.2$\%$ after $\sim$ 100 years. This shows that almost no water ice covers dust grains in Diffuse clouds, as supported by observations (\citealt{whittet2001}).

In translucent clouds, on the other hand, the medium is mostly molecular. We assume a density of 500 atoms~cm$^{-3}$, a visual extinction of 1 mag and an impinging UV radiation field of $G_0$ = 1. In this cloud, the deuterium and oxygen species are still in atomic form. Because in these clouds the hydrogen is in molecular form, the most efficient route to form H$_2$O is H$_2$ + O $\rightarrow$OH + H. At $T_{\rm{dust}}$ = 14~K, the hydrogen atoms evaporate, and water forms through the association of OH and H$_2$. Because the enthalpies of reactions are much lower for the formation of H$_2$O when the reaction involves H$_2$ than when it involves H atoms, the desorption \rm{upon formation}\rm\ becomes 3.2$\%$ for H$_2$ +O and 0.88$\%$ for H$_2$+OH. Therefore, the newly formed species mostly stay on the grain surface, and we find that no species are released into the gas phase ( see Fig.~\ref{env2}). The first layer of ice is growing until it reaches a coverage of few percent, confirming the absence of ice on the surface of dust grains in translucent clouds.

In dense clouds, the medium is also molecular, but denser. We consider a cloud  with a density of 5000 $cm^{-3}$, at a visual extinction of 5 mag, and an impinging UV field of $G_0$ = 1. At this extinction oxygen and deuterium are still in atomic form and the dust grains are colder, $T_{\rm{dust}}$ = 12~K. The formation of H$_2$O involves the reaction H$_2$ + O $\rightarrow$ OH + H. As discussed before, this reaction releases less energy than the reaction involving atomic H, and therefore only a small percentage of OH ($\sim$3~$\%$) is released into the gas phase. Because the dust grains are at lower temperature than in the  translucent cloud case, H atoms do not evaporate. The products of this reaction, OH and H, that stay on the surface can react and form H$_2$O. If these two products meet, 15~$\%$ of the formed H$_2$O is released into the gas phase. Our results, presented in Fig.~\ref{env3}, show that more than 10~$\%$ of the oxygen is released as H$_2$O into the gas phase. This means that the products H and OH have a probability of 60~$\%$ to meet each other. The OH that does not meet the H atoms will encounter an H$_2$ molecule and form H$_2$O + H. The part released into the gas phase through this process is negligible.

We also calculate the formation of molecules on dust grains in photo-dissociation regions. For this we use the results of \cite{meijerink2005} for a PDR with a density of 10$^3$ atoms $\rm{cm}^{-3}$, with a UV radiation field of G$_0$=10$^3$ and a visual extinction of 5 mag. Hydrogen is converted into H$_2$ but oxygen and deuterium are still in atomic form. The grain and gas temperatures are as high as 30~K. For these grain temperatures, the chemistry is very different than from above, as shown in Fig.~ref{env4}. H atoms as well as H$_2$ molecules, because of their low binding energy, evaporate efficiently, while oxygen atoms remain on the grain to form O$_2$ and O$_3$ molecules. The formation of water occurs through the successive hydrogenation of O$_2$ (\citealt{miyauchi2008}, \citealt{ioppolo2008}). H +O$_2$ $\rightarrow$ HO$_2$ and HO$_2$ + H $\rightarrow$ H$_2$O$_2$ and H$_2$O$_2$ + H $\rightarrow$ H$_2$O. Another important route is the hydrogenation of O$_3$, as shown by \cite{mokrane2009}: O$_3$+H $\rightarrow$ OH + O$_2$ and OH + H $\rightarrow$ H$_2$O.

\section{Discussion and conclusions}
We used Monte Carlo  methods in order to follow the chemistry involving H, D and O species on interstellar dust grain surfaces. The formation of surface species as well as species released into the gas phase is computed and allows us to estimate the efficiency for the formation of O bearing species. Our model shows a strong chemical differentiation with grain temperature. At $T_{\rm{dust}}$=10~K, processes involving hydrogenation are favored, while $T_{\rm{dust}}$=30~K favors oxygenation. UV photons can dissociate the species present on the surface, and the product of this dissociation can meet again and reform the initial species. Because lots of chemical reactions used in this work are exothermic, a fraction of the newly formed species desorbs directly into the gas phase. Therefore, the cycle formation-dissociation-formation increases the fraction of the species released into the gas phase significantly.  

We applied our model to different astrophysical environments. In diffuse clouds, water is formed through the successive hydrogenation of oxygen. These reactions are exothermic and allow an important fraction of newly formed OH and H$_2$O to desorb into the gas phase ($\sim$35 $\%$ and $\sim$15$\%$ respectively). In translucent clouds, on the other hand, the medium is molecular and the formation of water involves molecular hydrogen and atomic oxygen. These reactions produce a very small amount of energy, and therefore no species are released into the gas phase ($\sim$ 3$\%$ for OH and $\sim$ 1$\%$ for H$_2$O). In diffuse clouds, the temperature of the dust grains is slightly lower, and H atoms do not evaporate. Therefore, as in the case of translucents cloud, oxygen meets H$_2$ to give OH and H, but in this case the H atoms produced do stay on the grain. Water can form through the association of OH and this H (with a probability of 45$\%$), and again, an important fraction of H$_2$O is released into the gas phase ($\sim$7$\%$). 
In environments with warmer dust grains, such as PDRs, the chemistry occurring on the grain and the products released into the gas phase are very different. The formation of water involves O$_2$ (\citealt{ioppolo2008}) and O$_3$  (\citealt{mokrane2009}), and many oxygenated species are released into the gas phase.

\rm{The formation of ices requires very long time scales in diffuse and translucent clouds, as already discussed by \cite{cuppen2007}. In their work, these authors showed that a time of 10$^5$ years is required to form the first monolayer of ice in dense clouds, while water coverage would reach a maximum of 50~$\%$ of the surface in translucent clouds and 10~$\%$ of the surface in diffuse clouds. In our model, the water ice coverage reaches 0.3~$\%$ in diffuse clouds and 1~$\%$ in translucent clouds after 30 years.  We obtain a water ice coverage much lower than \cite{cuppen2007} since we consider that for each reaction that occurs on the surface, a certain fraction directly desorbs into the gas phase upon formation. In diffuse clouds, this fraction is of 36$\%$ and 15$\%$ for the formation of OH and H$_2$O respectively. Also, our model differs from \cite{cuppen2007} since it does not allow species to accrete on top of each other and therefore does not consider the formation of ices.}\rm

Even if our model does not allow us to estimate the coverage of ices and their morphology, the water coverage can be determined in steady state as an equilibrium between water formation and destruction by UV radiation. The formation rate of water on dust is set by R$_{\rm{H_2O}}$= n$_{\rm{O}}$ v$_{\rm{O}}$ $\rm{\sigma \epsilon}$, where n$_{\rm{O}}$ is the density of atomic oxygen in the gas phase, v$_{\rm{O}}$ the thermal velocity of oxygen atoms ($\sim$4.6 10$^4 \sqrt{\frac{T_{ \rm{gas}}}{100}} $cm s$^{-1}$) , $\sigma$ the cross section of the grain and $\epsilon$ the formation efficiency of water, that stays on the grain. \rm{ This efficiency directly depends on the formation route of water in different environments. In the case of diffuse clouds, water forms through the reactions O + H $\rightarrow$ OH, for which 64$\%$ stays on the dust grain (see section 2.3), and OH + H $\rightarrow$ H$_2$O for which 85$\%$ stays on the dust, yielding an efficiency on the order of 50~$\%$. In  translucent clouds,  water forms through O + H$_2$ $\rightarrow$ OH + H (96$\%$ stays on dust) and OH + H$_2$ $\rightarrow$ H$_2$O + H (99.2$\%$ stays on the dust), which brings the efficiency of water formation to $\sim$95$\%$. In dense clouds, this efficiency depends on the reactions  O + H$_2$ $\rightarrow$ OH + H and OH + H$\rightarrow$ H$_2$O, and therefore is on the order of $\sim$81$\%$. Note that in the PDR case, water formation follows another route involving O$_2$ that brings the efficiency to 5$\%$.}\rm\ The destruction by UV can be written as: R$_{\rm{des}}$=8~10$^{-10}$ G$_0$ $\rm{n_{H_2O}}$ $\exp({-2.2\rm{A_V}})$. In steady state, the fraction of the water molecules that covers the surface of a dust grain can be written as: 
\begin{equation}
\frac{\rm{n}_{H_2O}}{\rm{n_{site}}}=\frac{\rm{n_O v_O} \sigma_{\rm{site}} \epsilon}{8\times 10^{-10}G_0\exp({-2.2 A_V})}=1.725 10^{-6} \frac{\rm{n}_H \epsilon \sqrt{T_{\rm{gas}}}}{G_0 \exp({-2.2 A_V})}
\end{equation}
where  $\sigma_{\rm{site}}$ is the cross section of one site ($\sim$10$^{-15}$cm$^2$) and n$_{\rm{H}}$ the total density of hydrogen. This  fraction is $\sim$~0.3$\%$ in diffuse clouds, which is consistent with our calculations that show that this coverage is reached after 100 years (time to reach the steady state; see Fig.~\ref{env1}). In translucents clouds, the density is higher, and the coverage at steady state is of the order of 5~$\%$. For dense clouds, the medium is shielded, with an extinction of 5 magnitudes. The coverage becomes larger than 1, meaning that dust grains are covered by icy mantles with several layers. In PDR environments, with an high impinging UV radiation field (10$^3$), and for an extinction of 5 magnitudes, the coverage of ice is of 0.25~$\%$. These estimates, confirmed by our Monte Carlo simulations, show that ices that cover dust grains can only exist in shielded environments, if the UV radiation field is not too important. This is supported by observations of ices in diffuse and dense environments (\citealt{whittet2001}). \rm{\cite{jones1984}, estimate that the efficiency to form water on dust grains through repetitive hydrogenation of oxygen should be higher than 25$\%$ in order to reproduce the observations of the ice band in the Taurus molecular cloud. In this work, we find that the efficiency for the formation of water is at least of 50$\%$, which agrees with their findings. However, we have established that the successive hydrogenation of oxygen to form water was not important in molecular clouds, but that the reaction of oxygen with \hm\ is the dominant route. Indeed,  even if an important barrier exists for oxygen to react with \hm, these reactions are possible on surfaces since the species can repetitively meet each other which increases the cumulative probability to overcome the barrier. In this sense, dust grain surfaces can provide natural place for reactions with high barriers to occur, while these reactions would be highly unlikely in the gas phase.}\rm

A major result of this paper is that chemistry on bare grains has an important impact on gas phase chemistry because of the exo-thermicity of certain reactions on the surface. Once the grains become covered by icy mantles, the binding energies are much higher and the newly formed species remain trapped on the surface. Therefore, it would be very interesting to follow the chemistry that occurs in a diffuse cloud as it evolves into a dense cloud under the influence of hydrodynamics and gravity, and measure the impact of the grain surface chemistry on the gas phase composition while the grains are still bare grains. 

\rm We should mention that the probability for chemical species to desorb from grain surfaces upon formation has been seen experimentally (\citealt{pirronello1997}) and has been shown theoretically (\citealt{bachellerie2007}, \citealt{bergeron2008}). However, the dependance of this probability on the exo-thermicity of the reaction is unknown and has only been estimated empirically. The impact of grain surface chemistry on gas phase has been estimated in this work with the assumption that species forming on dust surfaces have higher probabilities to be released in the gas as the reaction is exothermic. Further studies are highly needed to assess this point and better understand the effect of grain surface chemistry on the gas phase composition. \rm

\clearpage
\begin{longtable}{ll|l|l|l}
\caption{\label{table1} Reactions adopted in our chemical model}
\\
{Reaction} & &Branching ratio&Barrier in K &references$^a$\\
\hline
H+ H &$\rightarrow$  H2&&0&\\
H+ D &$\rightarrow$  HD&&0&\\    
H+ O &$\rightarrow$  OH&&0&\\  
H+OH&$\rightarrow$  H2O&&0&\\
H+OD&$\rightarrow$  HDO&&0&\\
H+O2&$\rightarrow$  HO2&&250&\cite{walch1988}\\
H+H2O&$\rightarrow$  H2+OH&&9600&\\
H+HDO&$\rightarrow$  H2+OD&50$\%$&9600&\\
              &$\rightarrow$  HD+OH&50$\%$&9600&\\
H+D2O&$\rightarrow$  HD+OD&&9600&\\
H+O3&$\rightarrow$  O2 + OH&&450&\cite{lee1978} \\
H+HO2&$\rightarrow$  H2O2&&0&\\
H+DO2&$\rightarrow$  HDO2&&0&\\
H+H2O2&$\rightarrow$  OH + H2O&&$\sim$0&\cite{ioppolo2008}\\
H+HDO2&$\rightarrow$  OH + HDO&50$\%$&$\sim$0&\cite{ioppolo2008}\\
&$\rightarrow$  OD + H2O&50$\%$&$\sim$0&\cite{ioppolo2008}\\
H+D2O2&$\rightarrow$  OH + D2O&50$\%$&$\sim$0&\cite{ioppolo2008}\\
&$\rightarrow$  OD + HDO&50$\%$&$\sim$0&\cite{ioppolo2008}\\
D+ D &$\rightarrow$  D2&&0&\\   
D+ O &$\rightarrow$  OD&&0&\\ 
D+OH&$\rightarrow$  HDO&&0&\\
D+OD&$\rightarrow$  D2O&&0&\\
D+O2&$\rightarrow$  DO2&&250&\cite{walch1988}\\
D+H2O&$\rightarrow$  HD+OH&50$\%$&9600&\\
&$\rightarrow$H2+OD&50$\%$&9600&\\
D+HDO&$\rightarrow$  HD+OD&50$\%$&9600&\\
&$\rightarrow$  D2+OH&50$\%$&9600&\\
D+D2O&$\rightarrow$  D2+OD&&9600&\\
D+O3&$\rightarrow$  O2 + OD&&450&\cite{lee1978} \\
D+HO2&$\rightarrow$  HDO2&&0&\\
D+DO2&$\rightarrow$  OD + OD&&0&\\
D+H2O2&$\rightarrow$  OH + HDO&100$\%$&$\sim$0&\cite{ioppolo2008}\\
&$\rightarrow$  OD + H2O&0$\%$&$\sim$0&\cite{ioppolo2008}\\
D+HDO2&$\rightarrow$  OD + HDO&50$\%$&$\sim$0&\cite{ioppolo2008}\\
&$\rightarrow$  OH + D2O&50$\%$&$\sim$0&\cite{ioppolo2008}\\
D+D2O2&$\rightarrow$  OD + D2O&&$\sim$0&\cite{ioppolo2008}\\
O+H2 &$\rightarrow$  OH+H&&3000&\\   
O+HD &$\rightarrow$  OD+H&0$\%$$^b$&3000&\\   
 &$\rightarrow$  OH+D&100$\%$$^b$&3000&\\   
O+D2&$\rightarrow$  OD+D&&3000&\\   
O+ O &$\rightarrow$  O2&&0&\\ 
O+OH &$\rightarrow$  O2 + H&&0&\\
O+OD &$\rightarrow$  O2 + D&&0&\\
O+O2 &$\rightarrow$  O3&&0&\\
O+H2O &$\rightarrow$  OH + OH&&8500&\\
O+HDO &$\rightarrow$ OH + OD&&8500&\\
O+D2O &$\rightarrow$  OD + OD&&8500&\\
O+O3 &$\rightarrow$  O2 + O2&&2000&\\
O+HO2 &$\rightarrow$  OH + O2&&0&\\
O+DO2 &$\rightarrow$  OD + O2&&0&\\
O+H2O2 &$\rightarrow$  OH + HO2&&2000&\\
O+HDO2 &$\rightarrow$  OD + HO2&50$\%$&2000&\\
&$\rightarrow$  OH + DO2&50$\%$&2000&\\
O+D2O2 &$\rightarrow$  OD +D O2&&2000&\\
H2+ OH &$\rightarrow$  H2O+H&&2600&\cite{schiff1973}\\
H2+ OD &$\rightarrow$  HDO+H&100$\%$$^c$&2600&\cite{schiff1973}\\
&$\rightarrow$  H2O+D&0$\%$$^c$&2600&\cite{schiff1973}\\
H2+ O2 &$\rightarrow$  OH+OH&&35000& \\
H2+ HO2 &$\rightarrow$  H2O2+H&&10000& \\
H2+ DO2 &$\rightarrow$  HDO2+H&100$\%$&10000& \\ 
&$\rightarrow$  H2O2+D&0$\%$&10000& \\
HD+ OH &$\rightarrow$  HDO+H&50$\%$&2600&\cite{schiff1973}\\
&$\rightarrow$  H2O+D&50$\%$&2600&\cite{schiff1973}\\
HD+ OD &$\rightarrow$  HDO+D&50$\%$&2600&\cite{schiff1973}\\
&$\rightarrow$  D2O+H&50$\%$&2600&\cite{schiff1973}\\
HD+ O2 &$\rightarrow$  OH+OD&&35000& \\
HD+ HO2 &$\rightarrow$  HDO2 + H&50$\%$&10000& \\
&$\rightarrow$  H2O2 + D &50$\%$&10000& \\
HD+ DO2 &$\rightarrow$  D2O2+H&50$\%$&10000& \\
&$\rightarrow$  HDO2+D&50$\%$&10000& \\
D2+ OH &$\rightarrow$  HDO+D&100$\%$$^c$&2600&\cite{schiff1973}\\
&$\rightarrow$  D2O+H&0$\%$$^c$&2600&\cite{schiff1973}\\
D2+ OD &$\rightarrow$  D2O + D&&2600&\cite{schiff1973}\\
D2+ O2 &$\rightarrow$  OD + OD&&35000& \\
D2+ HO2 &$\rightarrow$  D2O2+H&0$\%$&10000& \\
&$\rightarrow$  HDO2+D&100$\%$&10000& \\
D2+ DO2 &$\rightarrow$  D2O2+D&&10000& \\
OH+ OH &$\rightarrow$  H2O2&&0&\\
OH+ OD &$\rightarrow$  HDO2&&0&\\
OH+ O2 &$\rightarrow$  HO2+O&&26000&\\
OH+ O3 &$\rightarrow$  HO2 + O2&&1000&\\
OH+ HO2 &$\rightarrow$  H2O + O2&&0&\\
OH+ DO2 &$\rightarrow$  HDO + O2&&0&\\
OH+ H2O2 &$\rightarrow$  H2O + HO2&&0&\\
OH+ HDO2 &$\rightarrow$  HDO + HO2&50$\%$&0&\\ 
&$\rightarrow$  H2O + DO2&50$\%$&0&\\
OH+ D2O2 &$\rightarrow$  HDO + DO2&100$\%$&0&\\
&$\rightarrow$  D2O + HO2&0$\%$&0&\\
OD+ O2 &$\rightarrow$  DO2+O&&26000&\\
OD+ O3 &$\rightarrow$  DO2 + O2&&1000&\\
OD+ HO2 &$\rightarrow$  HDO + O2&&0&\\
OD+ DO2 &$\rightarrow$  D2O + O2&&0&\\
OD+ H2O2 &$\rightarrow$  HDO + HO2&100$\%$&0&\\
 &$\rightarrow$  H2O + DO2&0$\%$&0&\\
 OD+ HDO2 &$\rightarrow$  HDO + DO2&50$\%$&0&\\ 
&$\rightarrow$  D2O + HO2&50$\%$&0&\\
OD+ D2O2 &$\rightarrow$  D2O + DO2&&0&\\
O2 + H2O2&$\rightarrow$  HO2 + HO2&&20000&\\
O2 + HDO2&$\rightarrow$  HO2 + DO2&&20000&\\
O2 + D2O2&$\rightarrow$  DO2 + DO2&&20000&\\
%
\multicolumn{5}{c}{$^a$ INIST Chemical Kinetics Database database; \cite{manion2008} if no ref. is indicated,$^b$\cite{robie1987},$^c$\cite{albers1971}}
\end{longtable}

\clearpage

\begin{table}
\caption{Physisorption and chemisorption energies of the considered species. \label{table2}}
\begin{tabular}{llll}
Species & Physisorption &Chemisorption \\
H&550&8500&\\
D&550&8500&\\    
O&1390&&\\
H2&600&&\\
HD&600&&\\    
D2&600&&\\
OH&1360&&\\
OD&1360&&\\    
O2&1440&&\\
H2O&2000&&\\
HDO&2000&&\\    
D2O&2000&&\\
O3&2900&&\\
HO2&2160&&\\
DO2&2160&&\\
H2O2&2240&&\\
HDO2&2240&&\\
D2O2&2240&&\\
\hline\hline
\end{tabular}
\end{table}

\begin{table}
\caption{Photodissociation reactions \label{phot2}}
\begin{tabular}{llll}
Reaction & $\alpha_{phot} (s^{-1})$&$\gamma$&CR$_0$$^d$\\
H2 + photon $\rightarrow$  H + H&5.0(-11)$^b$&a&0.05\\    
HD + photon $\rightarrow$  H + D&5.0(-11)$^b$&a&0.05\\    
D2 + photon $\rightarrow$  D + D&5.0(-11)$^b$&1.8&0.05\\    
OH + photon $\rightarrow$  O + H&3.9(-10)$^c$&2.24&254.5\\    
OD + photon $\rightarrow$  O + D&3.9(-10)$^c$&2.24&254.5\\    
H2O + photon$\rightarrow$  OH + H&8.0(-10)$^c$&2.2&485.5\\   
HDO + photon$\rightarrow$  OD + H&8.0(-10)$^c$&2.2&485.5\\   
D2O + photon$\rightarrow$  OD + D&8.0(-10)$^c$&2.2&485.5\\   
O2 + photon$\rightarrow$  O + O &7.9(-10)$^c$&2.13&375.5\\
O3 + photon$\rightarrow$  O2 + O &1.9(-9)$^c$&1.85&375.5\\
HO2 + photon $\rightarrow$  OH + O&6.7(-10)$^c$&2.12&375\\    
DO2 + photon $\rightarrow$  OD + O&6.7(-10)$^c$&2.12&375\\
H2O2 + photon $\rightarrow$  OH + O&9.5(-10)$^c$&2.28&750\\
HDO2 + photon $\rightarrow$  OD + O&9.5(-10)$^c$&2.28&750\\
D2O2 + photon $\rightarrow$  OH + O&9.5(-10)$^c$&2.28&750\\
\multicolumn{4}{l}{$\alpha_{\rm{phot}}=\alpha_{\rm{phot}}^0 \times \exp(-\gamma A_V)$ (s$^{-1}$);  
CR$_{\rm{phot}}$=10$^{-16}$ CR$_0$ (s$^{-1}$)} \\
\multicolumn{4}{c}{$^a$ depends on the H$_2$ and HD self-shielding; $^b$ van Dishoek (1998); $^c$ van Dishoek (2006); $^d$ Woodall et al. (2007)}
\end{tabular}
\end{table}

\begin{table}
\caption{Parameters used in our simulations \label{para}}
\begin{tabular}{l|l|l|l|l|l|l|l|l|l|}
Environment & n$_{\rm{H}}$ & n$_{\rm{HI}}$ & n$_{\rm{H_2}}$ & n$_{\rm{DI}}$ & n$_{\rm{OI}}$ & T$_{\rm{dust}}$ & T$_{\rm{gas}}$ & G$_0$ & A$_V$\\ \hline
Diffuse Clouds& 100 cm$^{-3}$&100 cm$^{-3}$& 0 cm$^{-3}$& 0.002 cm$^{-3}$ &0.03 cm$^{-3}$ &18~K&100~K&1&0.5 mag.\\
translucents clouds& 500 cm$^{-3}$&6 cm$^{-3}$& 247 cm$^{-3}$& 0.01 cm$^{-3}$& 0.15 cm$^{-3}$ & 14~K&50~K&1&1 mag.\\
Dense Clouds& 5000 cm$^{-3}$& 50 cm$^{-3}$& 2742 cm$^{-3}$& 0.1 cm$^{-3}$ & 1.5 cm$^{-3}$ &12~K&20~K&1& 5 mag.\\
PDR& 1000 cm$^{-3}$& 10  cm$^{-3}$& 495 cm$^{-3}$& 0.02 cm$^{-3}$ & 0.3 cm$^{-3}$ & 30~K & 30~K &10$^3$ & 5 mag\\
\multicolumn{4}{c}{}
\end{tabular}
\end{table}

\begin{figure}
\includegraphics[width=0.5\textwidth]{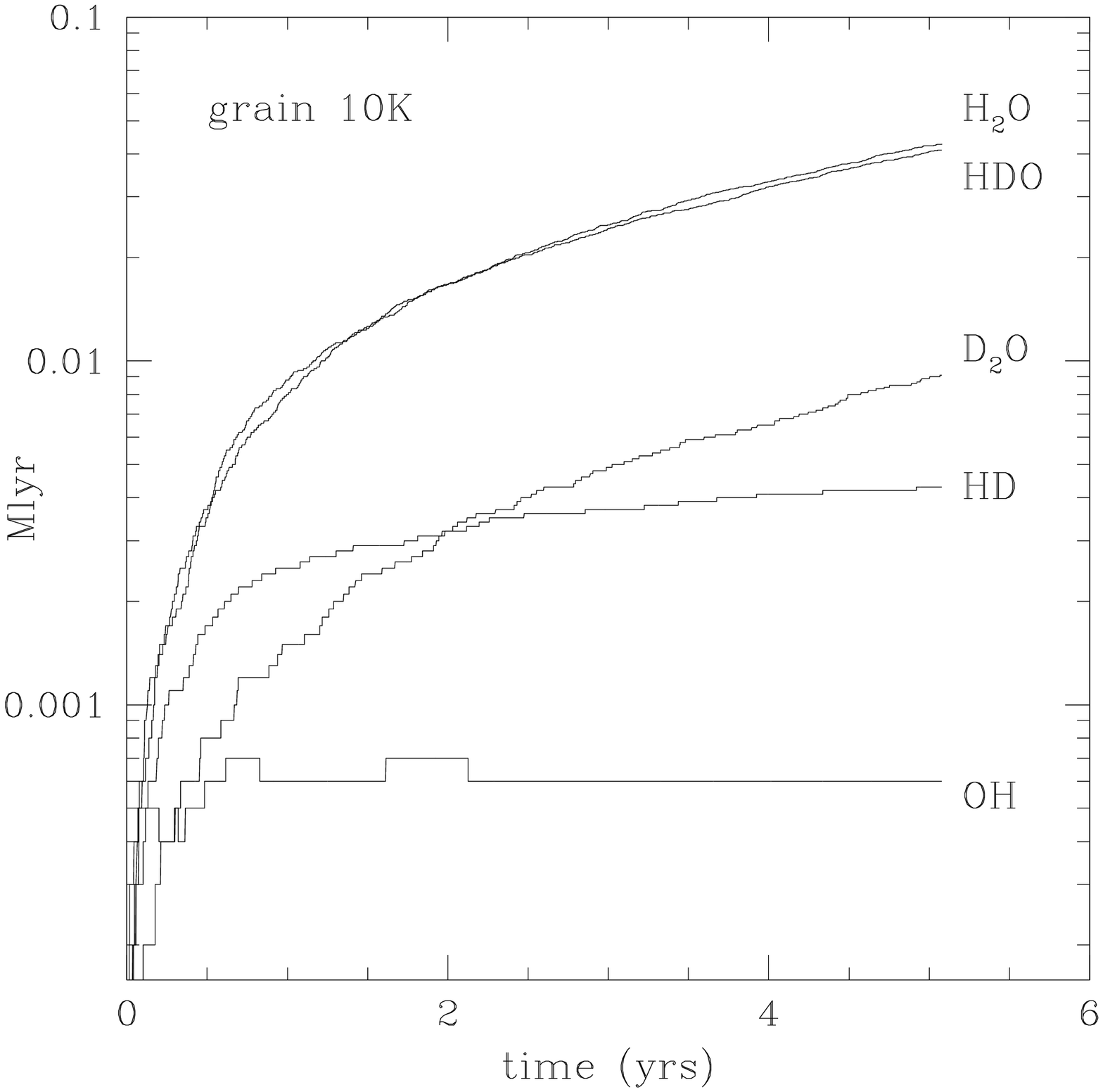}
\includegraphics[width=0.5\textwidth]{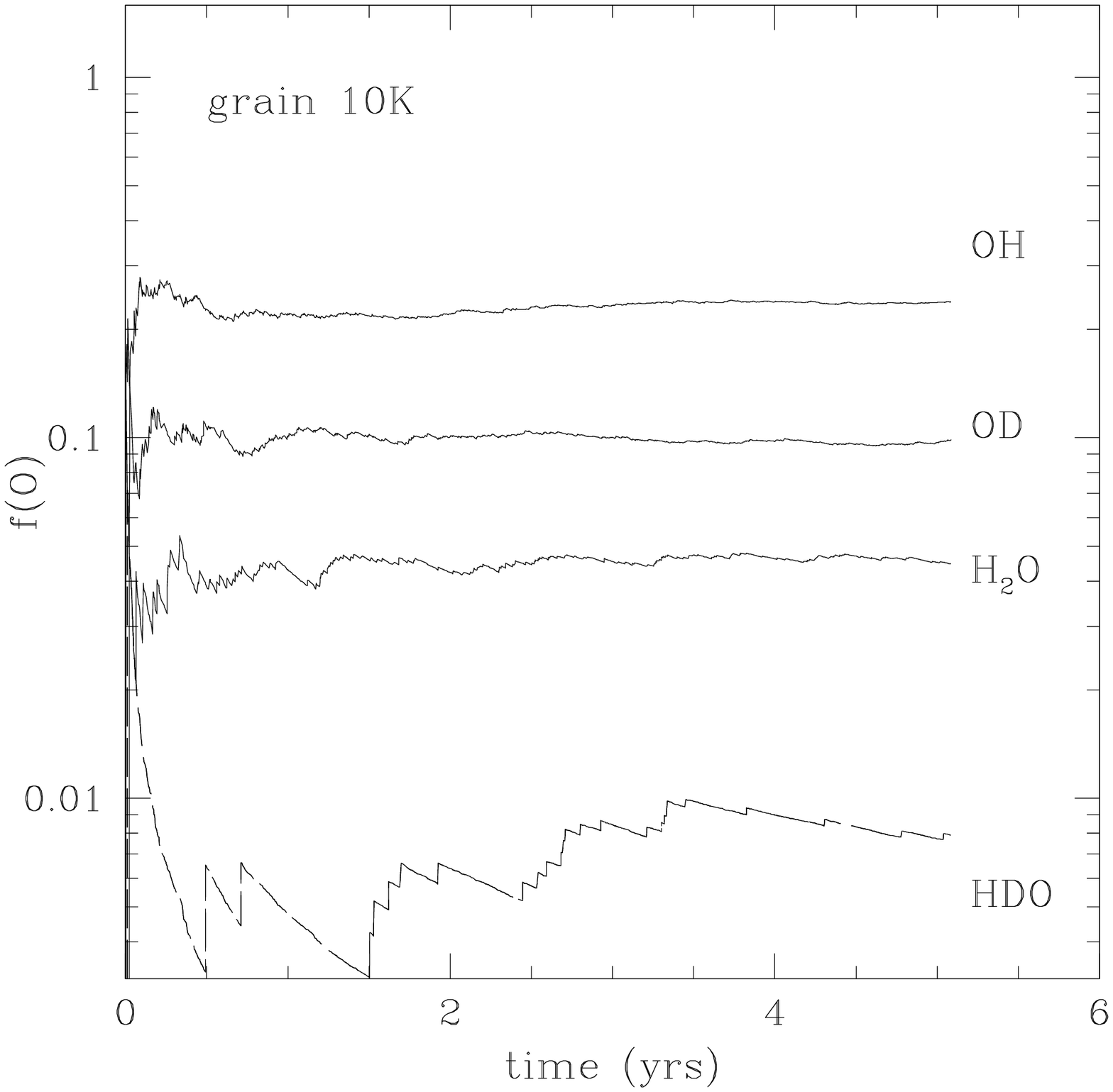}
\includegraphics[width=0.5\textwidth]{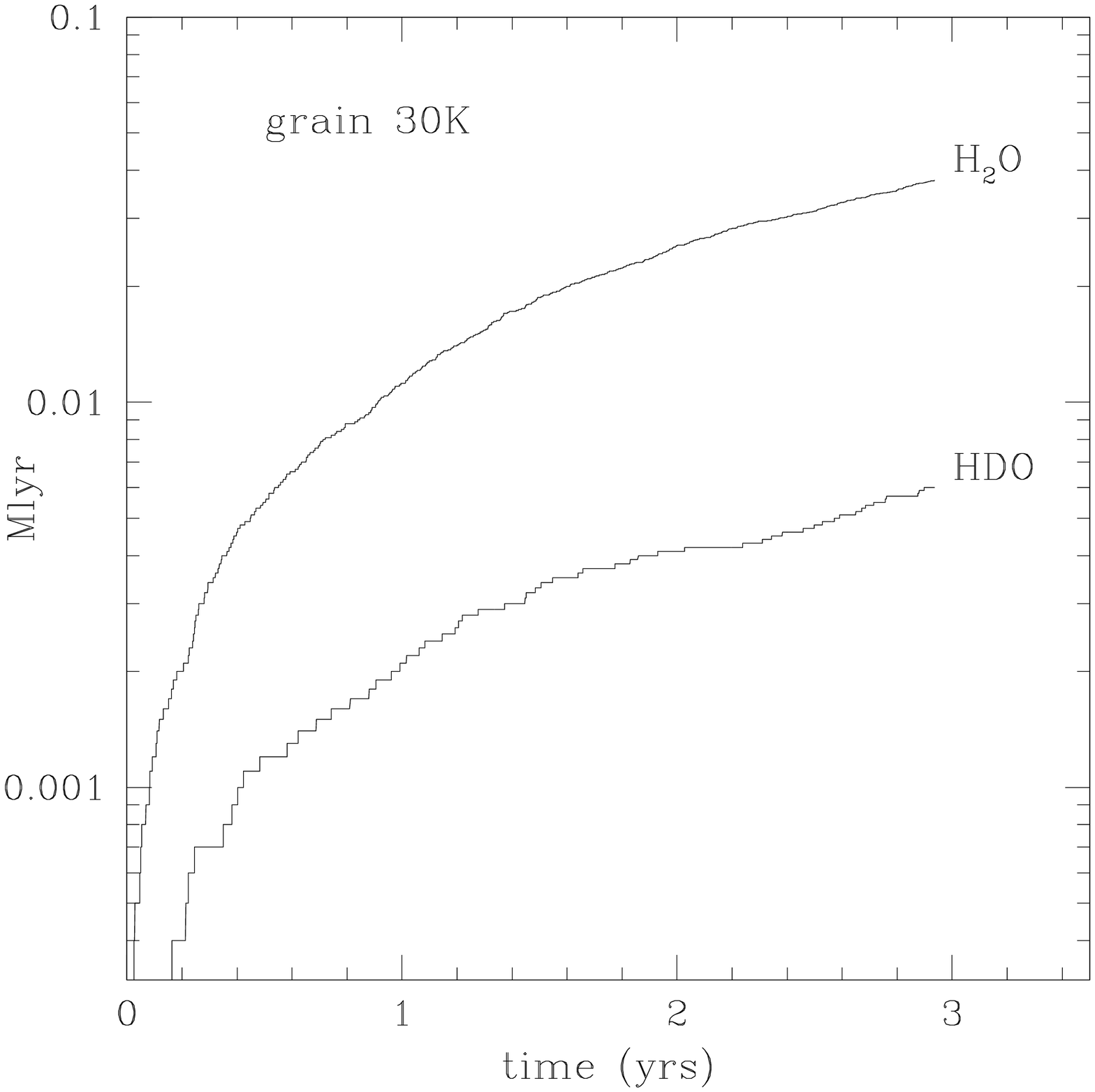}
\includegraphics[width=0.5\textwidth]{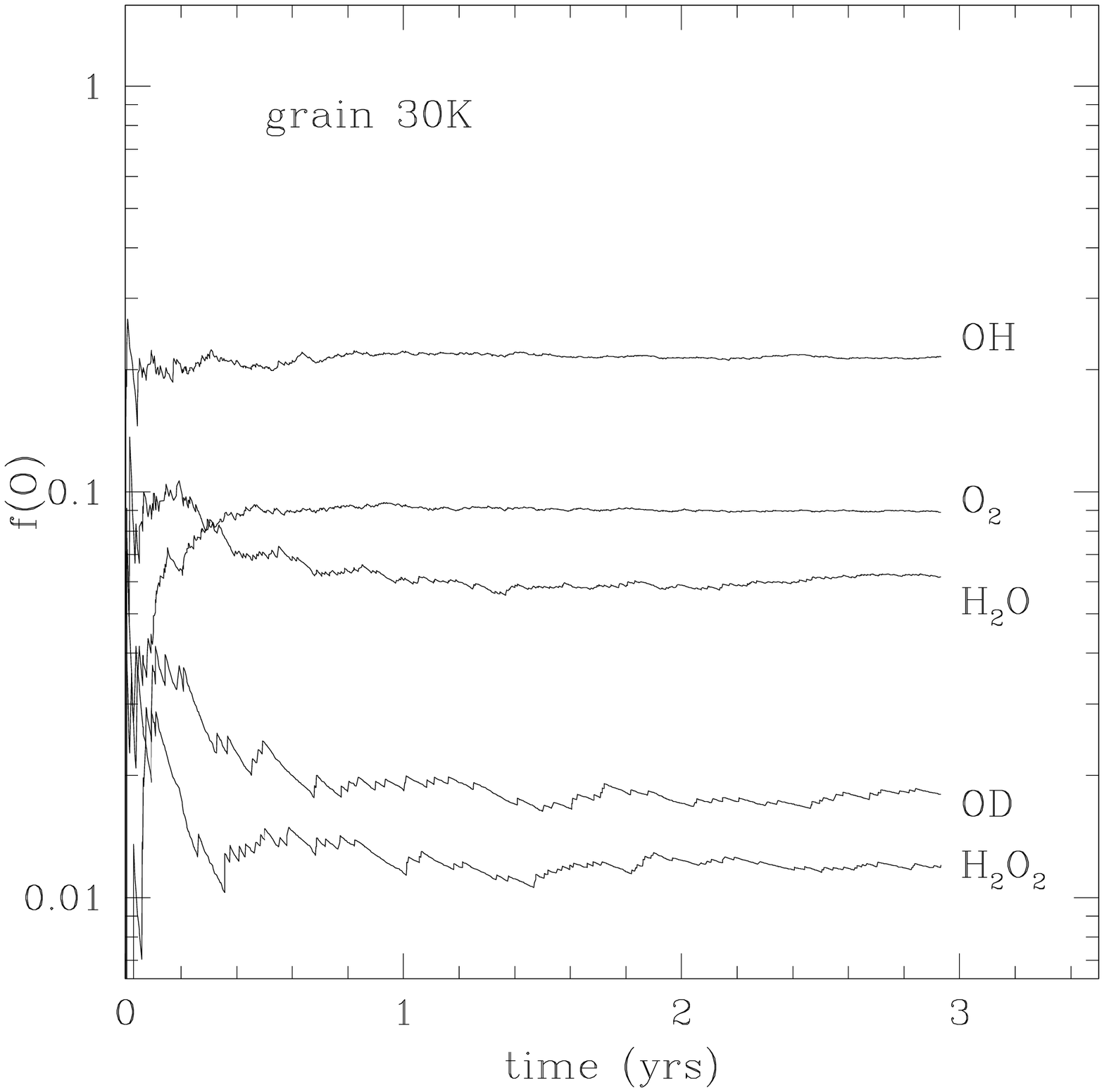}
\caption{Chemical species present on a grain surface of 10~K in monolayer (1 monolayer is a total coverage of the surface) as function of time (left panels). Fraction of newly formed oxygen bearing molecules f(O) that are released into the gas phase under the form of other species (right panels). Top: temperature of the grain surface is set to 10~K. Bottom: temperature of the grain surface is set to 30~K.  }
\label{temp1}
\end{figure}

\begin{figure}
\includegraphics[width=0.5\textwidth]{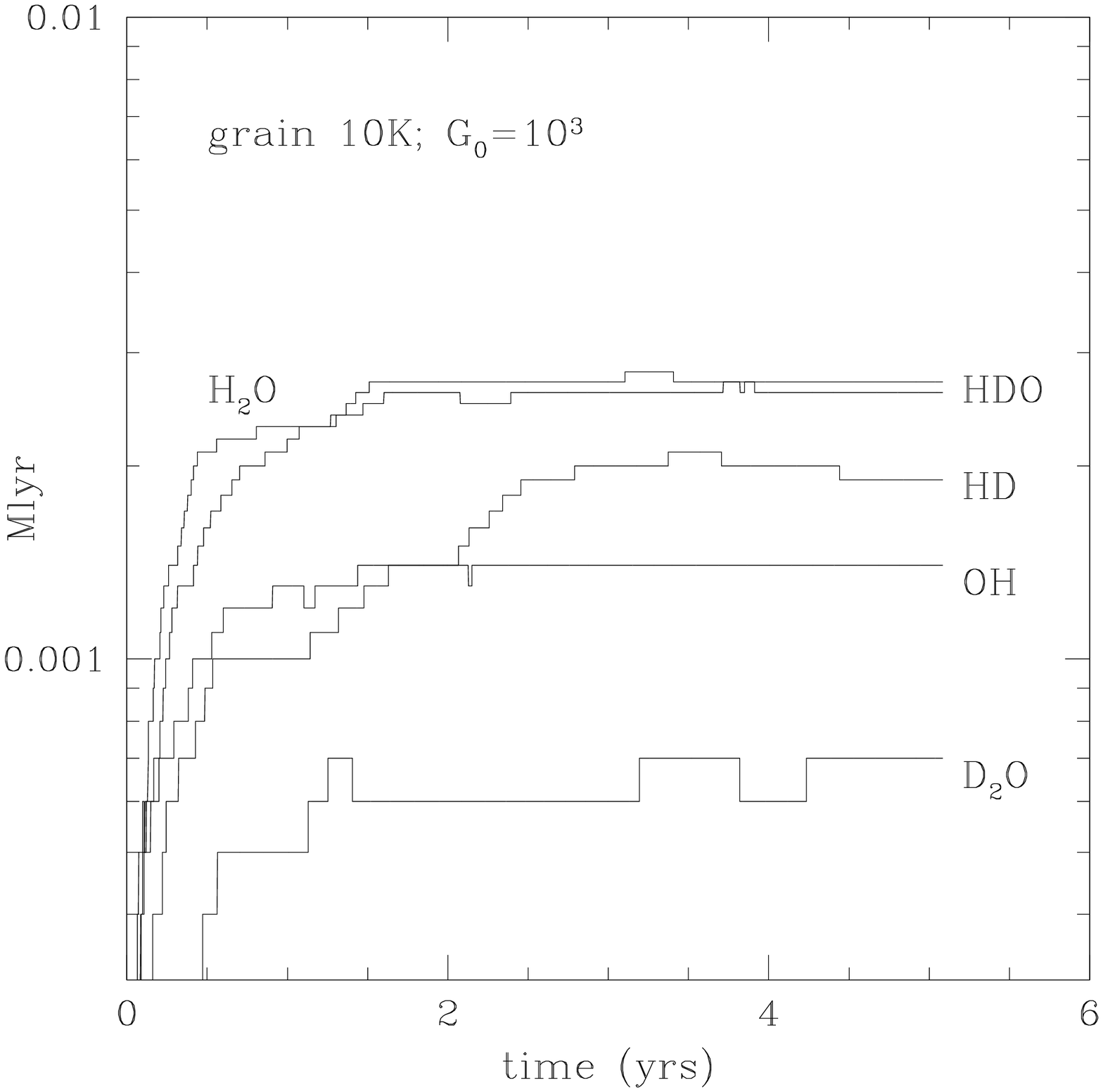}
\includegraphics[width=0.5\textwidth]{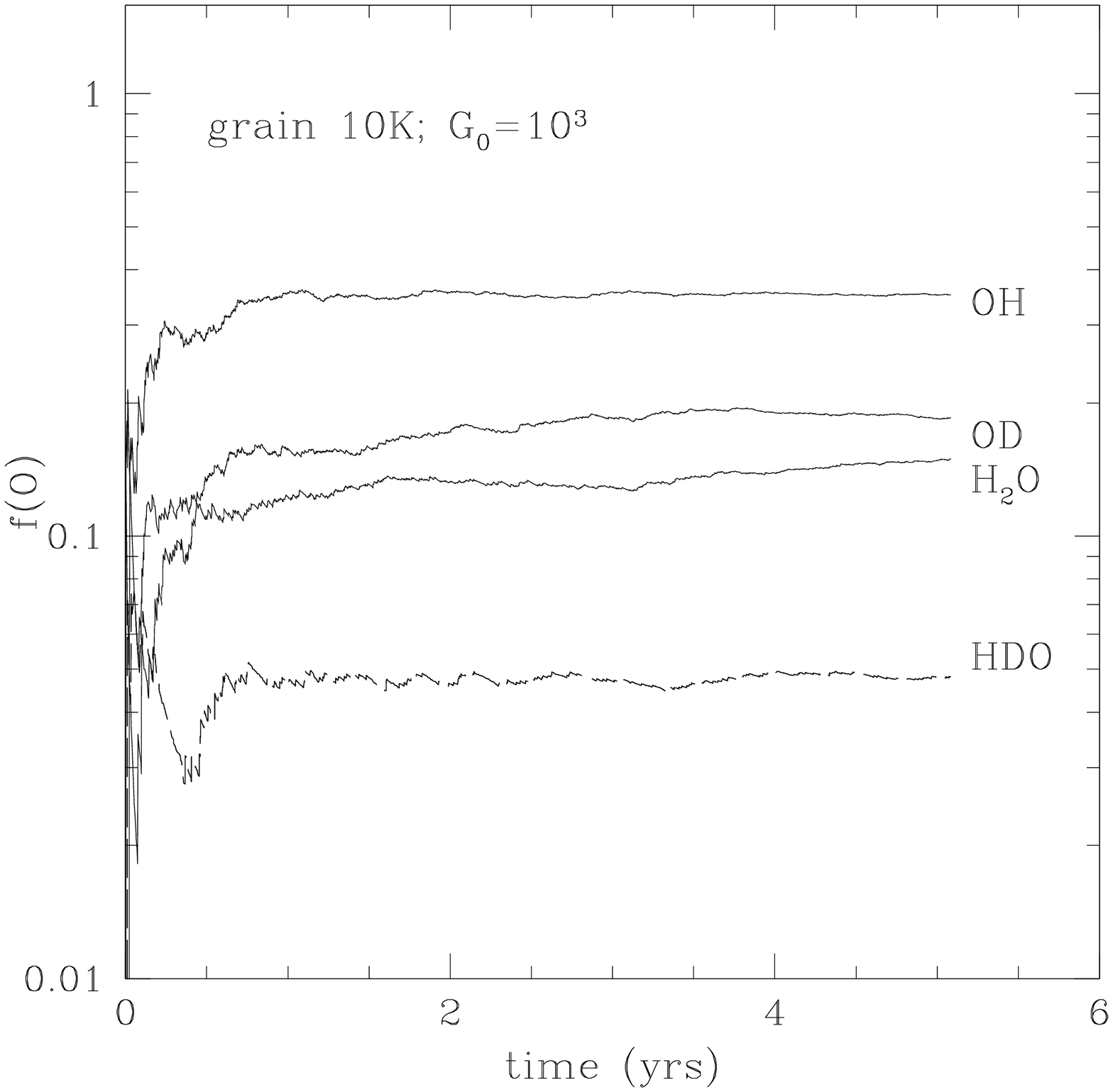}
\includegraphics[width=0.5\textwidth]{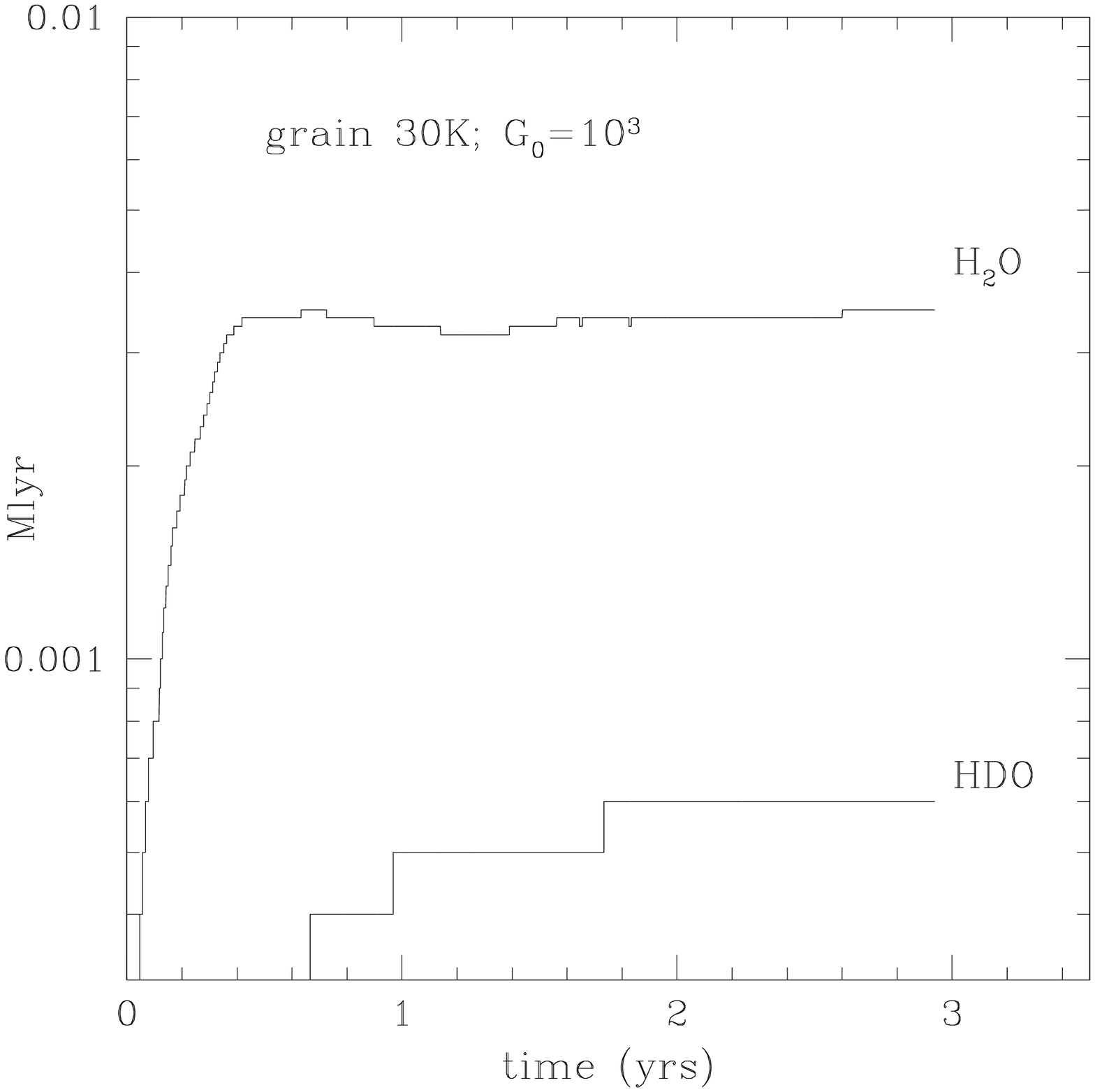}
\includegraphics[width=0.5\textwidth]{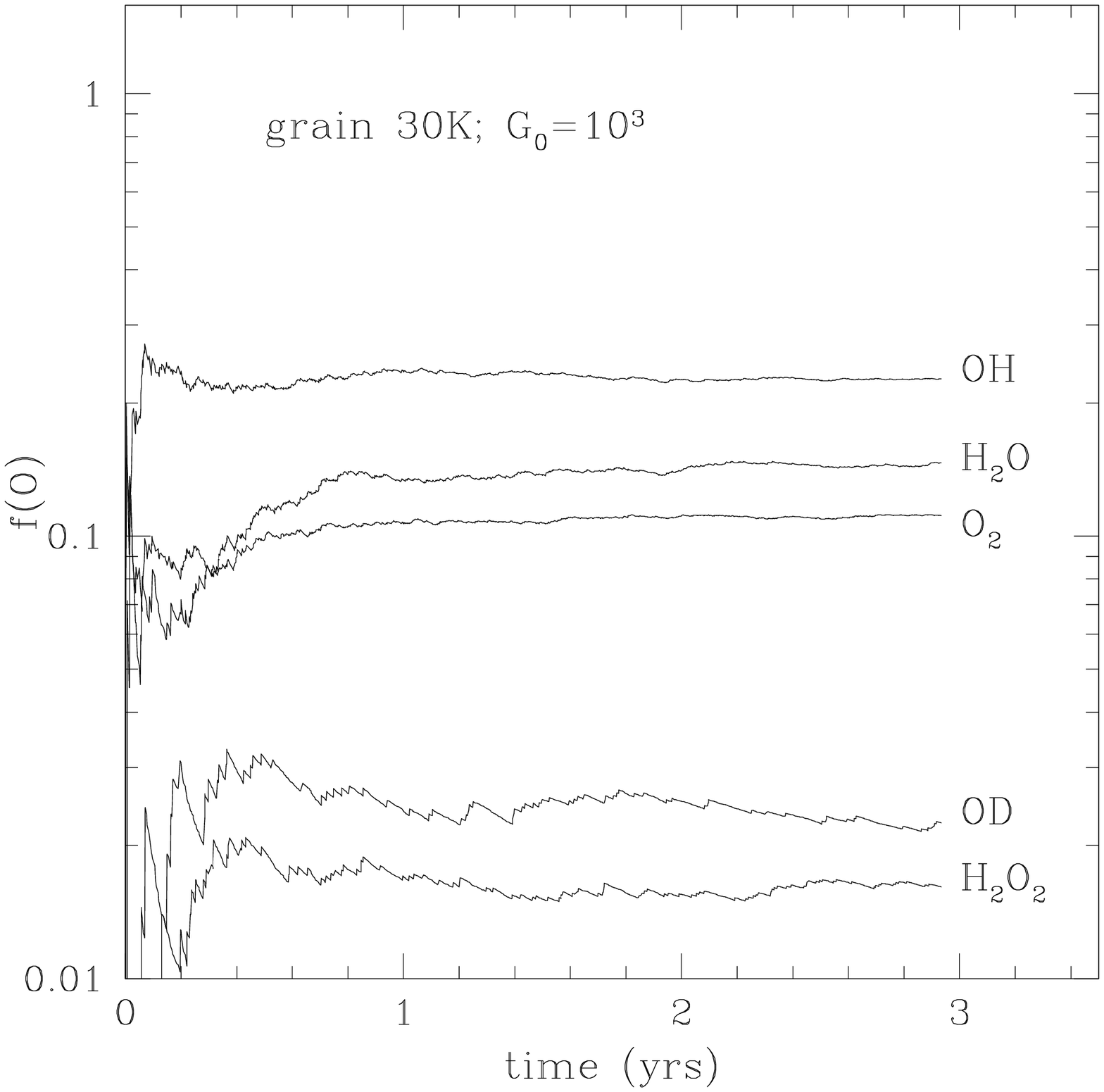}
\caption{Same as Fig.~\ref{temp1} with a UV radiation field of G$_0$=10$^3$. UV photons dissociate the chemical species present on the surface, which explains the low coverage on the grain surface (left panels). On the other hand, the amount of species released into the gas phase can be enhanced. }
\label{temp2}
\end{figure}

\begin{figure}
\includegraphics[width=0.5\textwidth]{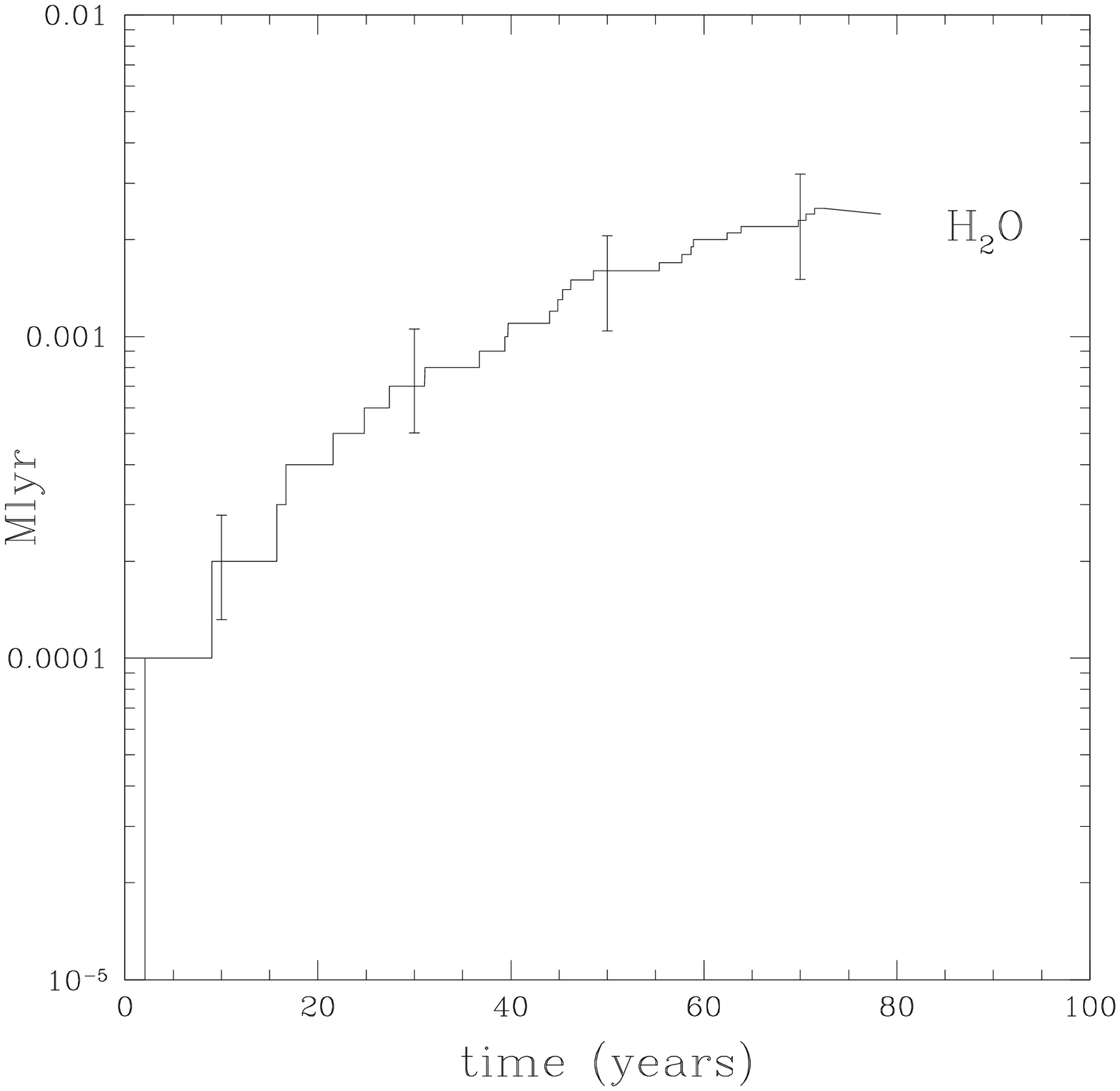}
\includegraphics[width=0.5\textwidth]{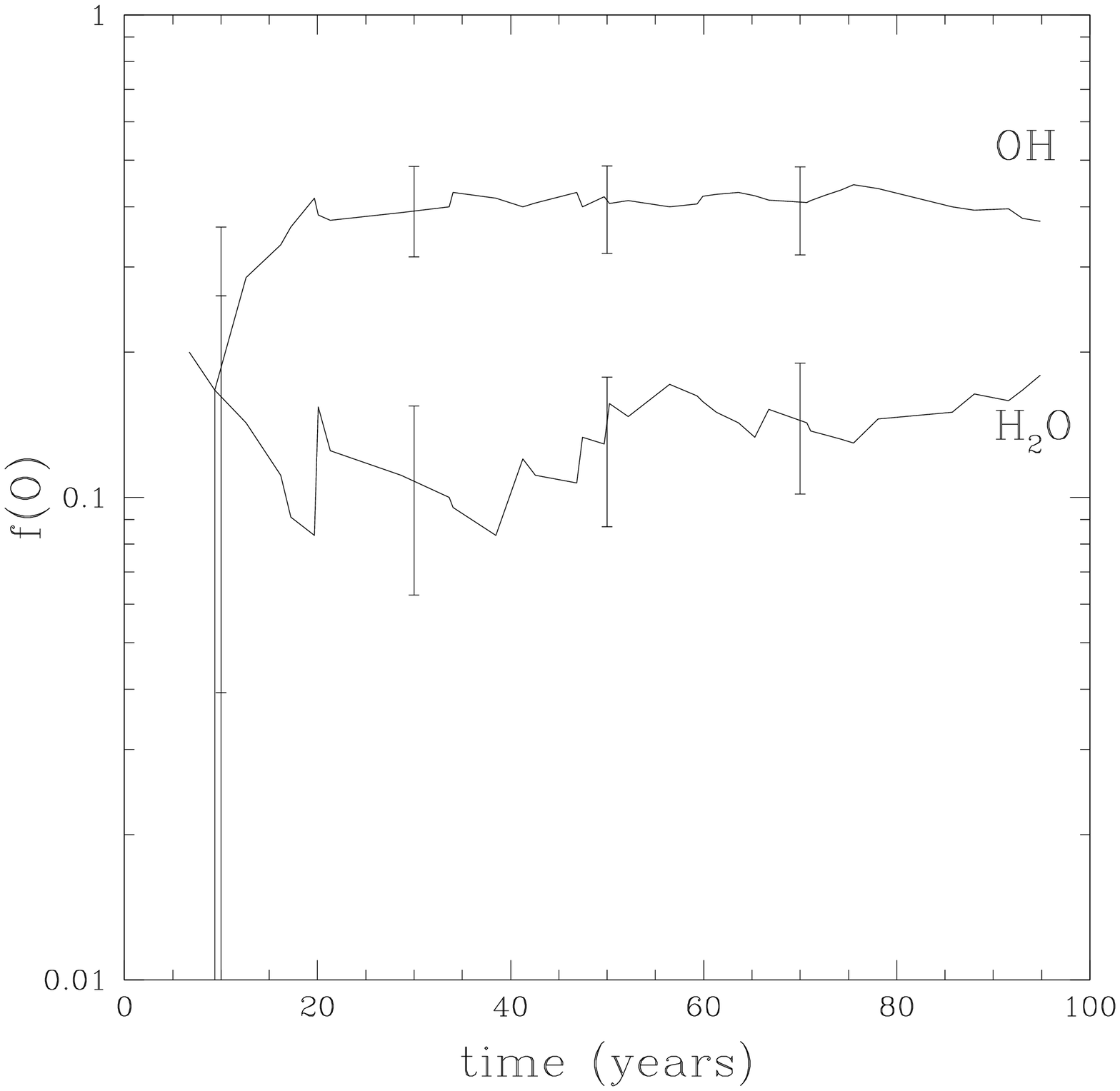}\\
\caption{Chemical species present on dust grains and released into the gas phase in diffuse clouds.  Left panel: chemical species present on a grain surface in monolayer (1 monolayer is a total coverage of the surface) as function of time. Right panel: fraction of newly formed oxygen bearing molecules f(O) that are released into the gas phase under the form of other species. \rm{The error bars  have been derived using 7 simulations and represent 95 $\%$ of confidence level.}\rm}
\label{env1}
\end{figure}

\begin{figure}
\includegraphics[width=0.5\textwidth]{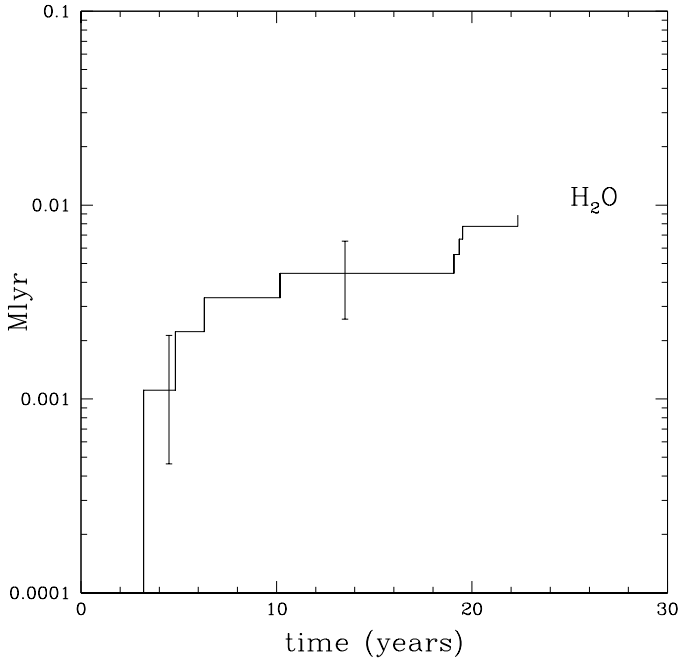}
\caption{Same as Fig.~\ref{env1} for translucent clouds. Note that no oxygen bearing species are released into the gas phase. \rm{The error bars  have been derived using 4 simulations and represent 95 $\%$ of confidence level.}\rm}
\label{env2}
\end{figure}

\begin{figure}
\includegraphics[width=0.5\textwidth]{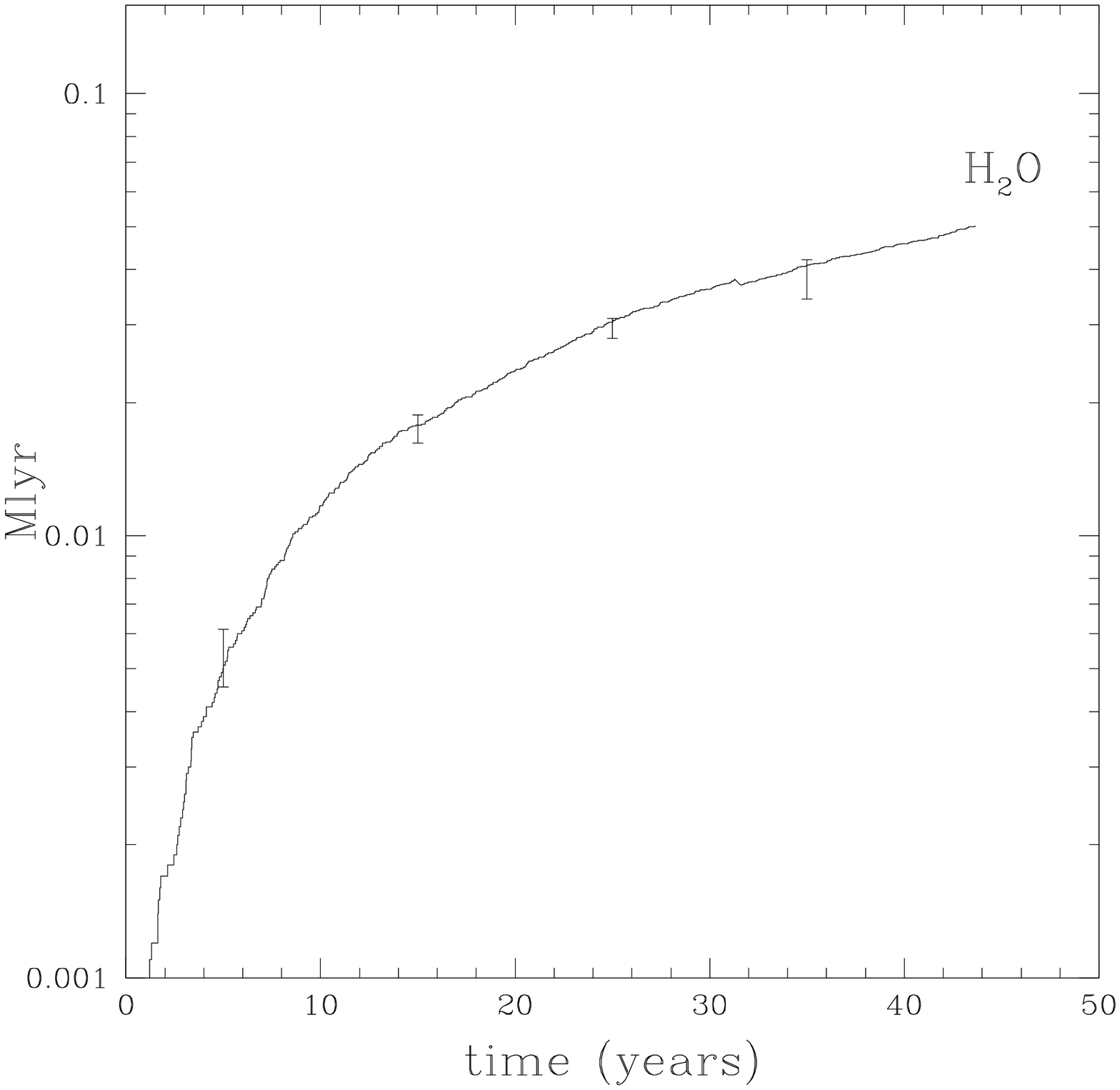}
\includegraphics[width=0.5\textwidth]{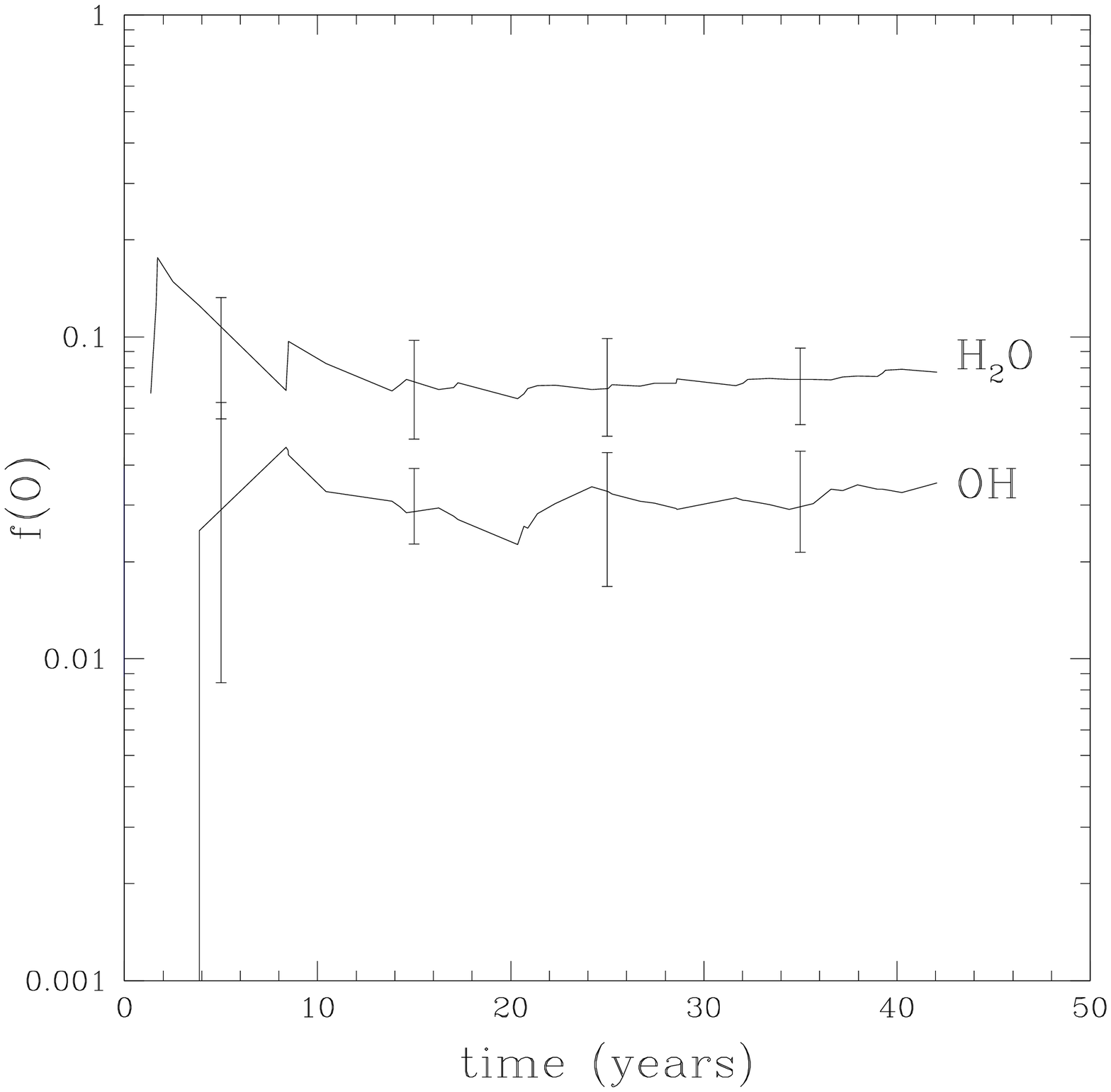}
\caption{Same as Fig.~\ref{env1} for dense clouds. \rm{The error bars  have been derived using 5 simulations and represent 95 $\%$ of confidence level.}\rm}
\label{env3}
\end{figure}

\begin{figure}
\includegraphics[width=0.5\textwidth]{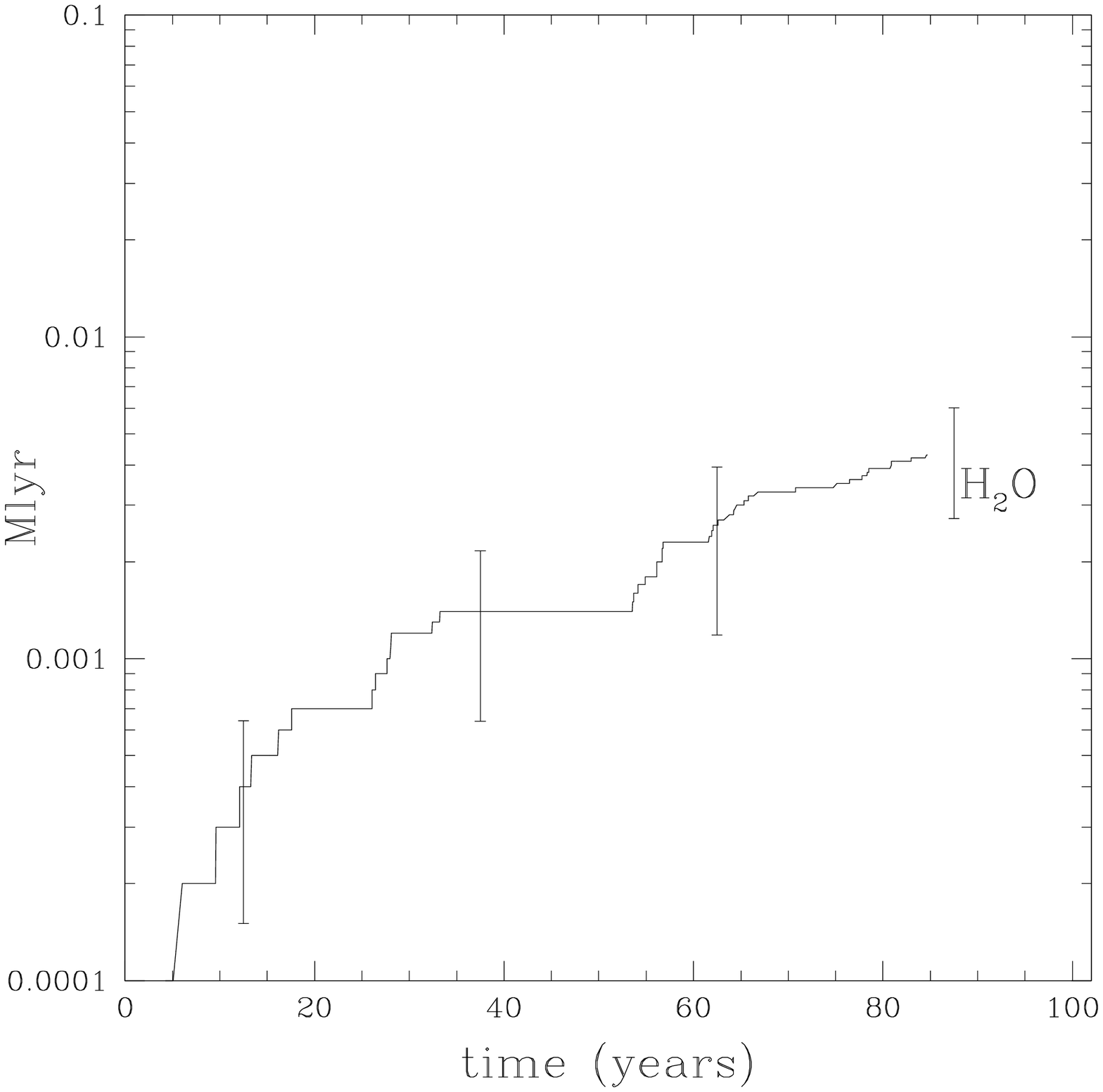}
\includegraphics[width=0.5\textwidth]{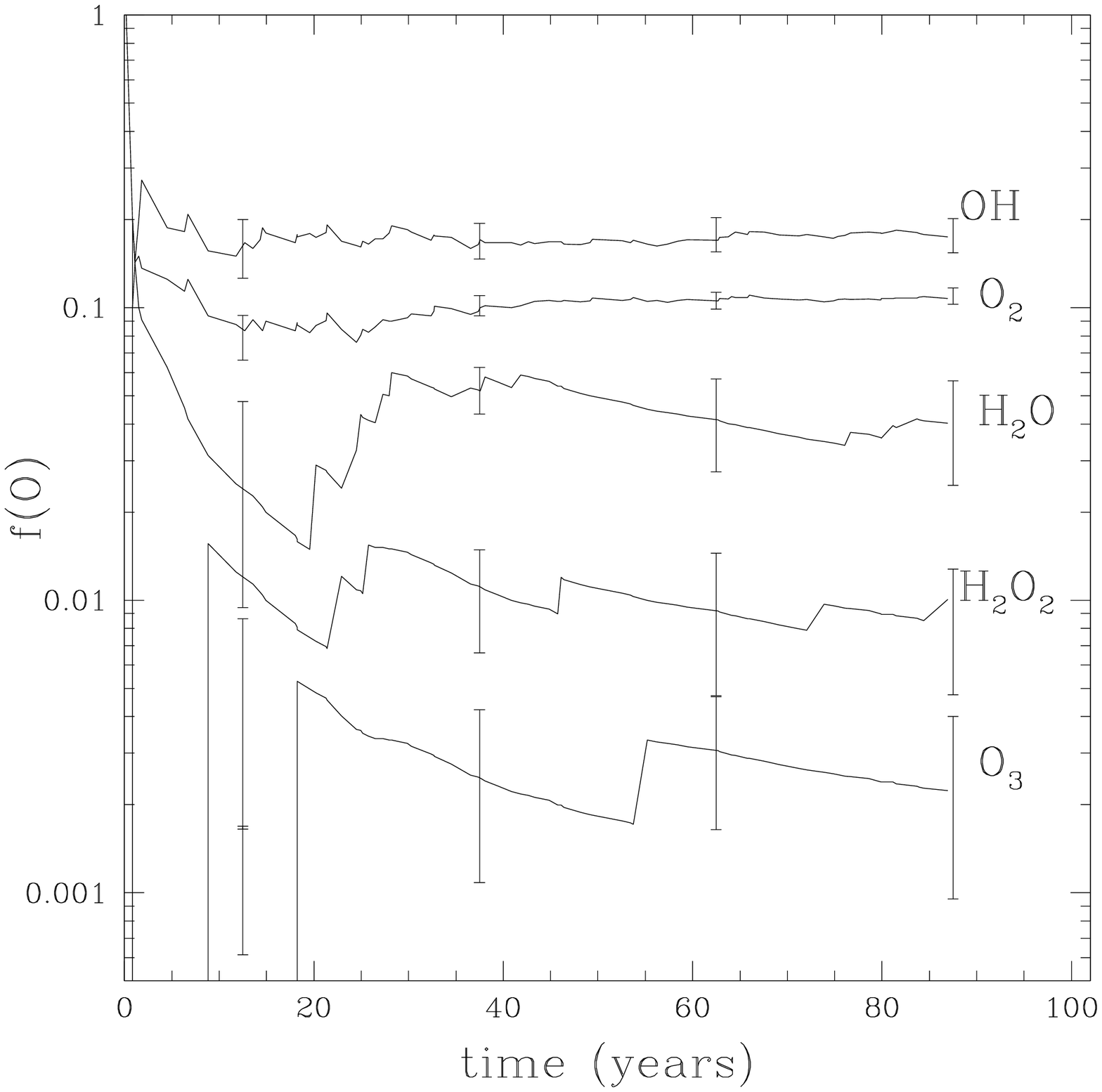}
\caption{Same as Fig.~\ref{env1} for PDRs. \rm{The error bars  have been derived using 7 simulations and represent 95 $\%$ of confidence level.}\rm}
\label{env4}
\end{figure}

\clearpage
\appendix
\section{Photo-desorption enhanced by photo-dissociation: recombination probability}

First, we assume that a photo-dissociation occurs, leading to one product on the $(x=0,y=0)$ site and the other next to it. The probability $P_{x,y}(i)$, at a given position $(x,y)$ on the grid, to find the product of the photo-dissociation after it has made $i$ steps can be found through previous probabilities at the step $i-1$:
\begin{equation}
P_{x,y} = \frac{1}{4}\left[P_{x-1,y}(i-1)+P_{x+1,y}(i-1)+P_{x,y-1}(i-1)+P_{x,y+1}(i-1)\right] 
\end{equation}
After further investigation, one can show that for the special position $(x=1,y=0)$, just next to the initial location, the probability can be written by separating even and odd cases:
\begin{eqnarray}
P_{1,0}(2i-1) & = & \frac{(i+1)^2}{4^{i+1}}\\
P_{1,0}(2i) & = & 0
\end{eqnarray}
More generally, the probablity $P^f_{x,y}(n)$ that the product arrives after exactly $n$ steps on a specified site situated at $(x,y)$ is given by:
\begin{equation}
P^f_{x,y}(n) = \left[\displaystyle\prod_{i=1}^{n-1}\left(1-P_{x,y}(i)\right)\right]P_{x,y}(n)
\end{equation}
If $n=2k,k\in\mathbb{N^*}$, then one can conclude that $P^f_{x,y}(n) = 0$, whereas if $n=2k-1,k\in\mathbb{N^*}$, $P^f_{x,y}(n)$ exists and can be simplified to:
\begin{equation}
P^f_{x,y}(2k-1) = \left[\displaystyle\prod_{j=1}^{k-1}\left(1-P_{x,y}(2j-1)\right)\right]P_{x,y}(2k-1).
\end{equation}
Hence for the case where $x=1$ and $y=0$:
\begin{equation}
P^f_{1,0}(2k-1) = \left[\displaystyle\prod_{i=2}^k\left(1-4^{-i}i^2\right)\right]\frac{(k+1)^2}{4^{k+1}}. 
\end{equation}
In this last equation the product term can be seen as a function of integers $\varphi(k)$ having for first terms:
\begin{eqnarray}
\varphi(k) & = & \displaystyle\prod_{i=2}^k 1-4^{-i}i^2\\
\varphi(1) & = & 1 \\
\varphi(2) & = & 0.75 \\
\varphi(3) & \simeq & 0.6445 \\
\varphi(4) & \simeq & 0.6042 \\
\displaystyle\lim_{k\rightarrow +\infty}\varphi(k) & \simeq & 0.581739
\end{eqnarray}
Here we see how fast this function converges to an unique value. At the end, the probability $P_{1,0}$ that the initial product comes back to its site of production is then:
\begin{equation}
P_{1,0} = \displaystyle\sum_{i=1}^{\infty}S(i)P^f_{1,0}(i),
\end{equation}
where $S(i)$ is the cumulative distribution of the number of steps that the product will do. According to the previous results that we obtained, this probability can be re-written as:
\begin{equation}
P_{1,0} = \displaystyle\sum_{k=1}^{\infty}S(2k-1)\varphi(k)\frac{(k+1)^2}{4^{k+1}}.
\end{equation}
First, if we assume that the product can make as many steps as it wants ($\forall k, S(2k-1)=1$), $P_{1,0}$ can be approximated as:
\begin{equation}
P_{1,0} \simeq \frac{1}{4}\varphi(1)+\frac{9}{64}\varphi(2)+\frac{25}{256}\varphi(3)+\frac{3}{85}\varphi(4)+\varphi^*\displaystyle\sum_{k=5}^{+\infty}\frac{(k+1)^2}{4^{k+1}}
\end{equation}
where $\varphi^*$ is the limit value when $k\rightarrow +\infty$. To have a better idea of how fast this serie converges, note that the sum in the last term of the previous equation has an algebraically defined value which is very small compared to the other terms. Indeed:
\begin{equation}
\varphi^*\displaystyle\sum_{k=5}^{+\infty}\frac{(k+1)^2}{4^{k+1}} = 0.581739\times\frac{365}{27648} \simeq 0.00768.
\end{equation}
Here we see that the direct recombination effect is enhanced mainly by the first term of the sum, \textit{i.e.} when $k<5$, hence a
number of steps smaller than $9$. This shows that one can assume that a product has left its initial position after 10 steps if it has
not come back at least one time to its initial position (or neighbourhood here). Finally, we can estimate $P_{1,0}$ to:
\begin{equation}
P_{1,0} \simeq 0.4474,
\end{equation}
giving a new intrinsic probability of photo-desorption $P_\mathrm{spd}$ enhanced compared to the initial value. Indeed, if we take into account that a new molecule that is not photo-desorbed can be splitted by photo-dissociation, then the product can encounter for a second time the reactant, giving him a second chance to photo-desorb, and so on. Thus, an enhanced value of the spontaneous photo-desorption $P^*_\mathrm{spd}$ can be expressed as:
\begin{eqnarray}
P^*_\mathrm{spd} & = & P_\mathrm{spd} + (1-P_\mathrm{spd})P_{1,0}P_\mathrm{spd}+(1-P_\mathrm{spd}P_{1,0})^2 P_\mathrm{spd} + ... \\
 & = & P_\mathrm{spd}\displaystyle\sum_{i=0}^{+\infty}\left[(1-P_\mathrm{spd})P_{1,0}\right]^i.
\end{eqnarray}
The last term being a geometric series that converges because $|(1-P_\mathrm{spd})P_{1,0}|<1$, and one can simplify to:
\begin{equation}
P^*_\mathrm{spd} = \frac{P_\mathrm{spd}}{1-(1-P_\mathrm{spd})P_{1,0}}.
\end{equation} 
In the case of water formation with the reaction O+ H $\rightarrow$ OH and OH + H  $\rightarrow$ H$_2$O, the fraction of water that desorbs upon formation is  $5\%$. In the presence of a UV field, this value becomes  $9.4\%$.

\end{document}